\documentclass[a4paper,11pt]{article}
\pdfoutput=1 

\usepackage{jheppub} 
\usepackage{amsmath,amssymb,amsthm,amscd,graphicx}
\input epsf.sty

\theoremstyle{definition}

\newcommand{\de}{\partial}

\newcommand{\CC}{{\cal C}}

\newcommand{\CN}{{\cal N}}
\newcommand{\CO}{{\cal O}}

\def\IR{{\mathbb R}}
\def\IC{{\mathbb C}}

\def\IP{{\mathbb P}}

\def\IF{{\mathbb F}}

\newcommand{\tr}{{\rm Tr}}
\newcommand{\re}{{\rm e}}
\newcommand{\ri}{{\rm i}}
\newcommand{\rd}{{\rm d}}

\newcommand{\mJ}{\mathsf{J}}

\newcommand{\mO}{\mathsf{O}}

\newcommand{\mS}{\mathsf{S}}

\newcommand{\mx}{\mathsf{x}}

\newcommand{\mU}{\mathsf{U}}

\newcommand{\mz}{\mathsf{z}}
\newcommand{\mm}{\mathsf{p}}

\newcommand{\mH}{\mathsf{H}}

\newcommand{\be}{\begin{equation}}
\newcommand{\ee}{\end{equation}}
\newcommand{\ba}{\begin{aligned}}
\newcommand{\ea}{\end{aligned}}
\newcommand{\ben}{\begin{eqnarray}\displaystyle}
\newcommand{\een}{\end{eqnarray}}

\newcommand{\sectiono}[1]{\section{#1}\setcounter{equation}{0}}

\newdimen\tableauside\tableauside=1.0ex
\newdimen\tableaurule\tableaurule=0.4pt
\newdimen\tableaustep
\def\phantomhrule#1{\hbox{\vbox to0pt{\hrule height\tableaurule width#1\vss}}}
\def\phantomvrule#1{\vbox{\hbox to0pt{\vrule width\tableaurule height#1\hss}}}
\def\sqr{\vbox{%
  \phantomhrule\tableaustep
  \hbox{\phantomvrule\tableaustep\kern\tableaustep\phantomvrule\tableaustep}%
  \hbox{\vbox{\phantomhrule\tableauside}\kern-\tableaurule}}}
\def\squares#1{\hbox{\count0=#1\noindent\loop\sqr
  \advance\count0 by-1 \ifnum\count0>0\repeat}}
\def\tableau#1{\vcenter{\offinterlineskip
  \tableaustep=\tableauside\advance\tableaustep by-\tableaurule
  \kern\normallineskip\hbox
    {\kern\normallineskip\vbox
      {\gettableau#1 0 }%
     \kern\normallineskip\kern\tableaurule}%
  \kern\normallineskip\kern\tableaurule}}
\def\gettableau#1{\ifnum#1=0\let\next=\null\else
\squares{#1}\let\next=\gettableau\fi\next}

\newcommand{\figref}[1]{Fig.~\protect\ref{#1}}

\title{\huge{Resonances and PT symmetry in quantum curves}}

\author[a]{Yoan Emery,}
\affiliation[a]{D\'epartement de Physique Th\'eorique et Section de Math\'ematiques\\
Universit\'e de Gen\`eve, Gen\`eve, CH-1211 Switzerland}
\emailAdd{yoan.emery@unige.ch}

\author[a]{Marcos Mari\~no,}
\emailAdd{marcos.marino@unige.ch} 

\author[a,b]{Massimiliano Ronzani\,}
\affiliation[b]{Scuola Internazionale Superiore di Studi Avanzati (SISSA) \\via Bonomea 265, 34136 Trieste, Italy}
\emailAdd{ronzani.massimiliano@gmail.com}

\abstract{In the correspondence between spectral problems and topological strings, it is natural to consider complex values for the string theory moduli. 
In the spectral theory side, this corresponds to non-Hermitian quantum curves with complex spectra and resonances, and in some cases, to PT-symmetric spectral problems. The correspondence leads to precise predictions about the spectral properties of these non-Hermitian operators. In this paper we 
develop techniques to compute the complex spectra of these quantum curves, providing in this way precision tests of these predictions. 
In addition, we analyze quantum Seiberg--Witten curves with PT symmetry, which provide interesting and exactly solvable examples of spontaneous PT-symmetry breaking.}    

\begin{document}
\maketitle
\flushbottom


\sectiono{Introduction}

In recent years, various dualities between string/gravity theories and gauge theories have been proposed, in which the spacetime in the gravity side emerges from 
the strong coupling dynamics of the quantum gauge theory. In these dualities, a precise dictionary relates 
the geometric moduli of the spacetime to global parameters of the quantum theory. 
One interesting and somewhat puzzling aspect of this dictionary is that their entries have sometimes a very different nature. For example, the 
gauge theory side involves 
global parameters which are {\it discrete} and {\it real} (like for example the rank of the gauge group). These parameters correspond, in the string theory side, 
to {\it continuous} moduli which often can take {\it complex} values. In order to 
better understand these dualities, it is important to address these hidden tensions in the dictionary. 

Dualities involving topological strings have provided simplified models in which many of these questions 
can be addressed in great detail. In \cite{ghm,cgm}, a precise correspondence was 
postulated between topological string theory on toric Calabi--Yau (CY) manifolds, and 
quantum-mechanical systems on the real line. This correspondence, sometimes referred to as TS/ST 
correspondence, leads to highly non-trivial mathematical predictions, and can be used as an 
excellent testing ground to examine the subtleties involved in the dictionary relating string 
theory to quantum systems. 

In this paper, we will address the tension between real variables in the quantum theory side, and complex variables 
in the geometry side\footnote{Some aspects of the tension between discrete and continuous variables were addressed in \cite{cgm8}.}. In 
the TS/ST correspondence, some of the K\"ahler moduli of the CY correspond to parameters in the Hamiltonian. 
On the topological string side, K\"ahler moduli are naturally complex (in fact, they are mapped to complex structure moduli through mirror symmetry). 
However, on the spectral theory side, complex parameters lead to non-Hermitian operators. In general, the spectral theory of non-Hermitian operators 
is much more subtle than the one of Hermitian operators, and a precise definition of the spectral problem is needed. Usually, 
one has to perform an analytic continuation of the eigenvalue problem as the parameter under consideration rotates into the complex plane. This often 
uncovers a rich structure, as was demonstrated in the pioneering work of Bender and Wu on the quartic oscillator \cite{bw}. An important 
example of such a situation occurs when there are {\it resonances}: in this case, the operator is symmetric but it does not have conventional 
bound states, nor a real spectrum. For example, in non-relativistic quantum mechanics,  
real potentials which are not bounded from below, like the cubic oscillator or the inverted quartic oscillator, lead to a resonant spectrum of complex eigenvalues. 

For the operators appearing in the TS/ST correspondence, the non-Hermitian case is largely unexplored, since one has to deal with difference operators.  
We are then in a situation in which one aspect of the theory -its behavior when the parameters become complex- is easy to understand on one side of the duality 
(in this case, on the topological string theory side), but is more difficult on the other side (in this case, in the spectral theory side). One can then 
use the topological string to obtain predictions for the spectral theory problem in the non-Hermitian case. One such prediction, already noted in 
\cite{cgum}, is that we will have generically an infinite discrete spectrum 
of {\it complex} eigenvalues, which can be easily computed on the string theory side by analytic continuation. The analysis of \cite{cgum} focused on 
real values of the parameters leading to resonances. However, the predictions from topological string theory were not verified on the spectral theory side\footnote{In this paper we consider the case 
in which the mass parameters become complex, while $\hbar$ remains real. The case in which $\hbar$ is complex was studied in \cite{gm17,kaserg}.}. 

The first goal of this paper is to develop methods to compute this spectrum of complex 
eigenvalues, directly in spectral theory. In conventional quantum mechanics there are two different methods to compute such a spectrum. 
The first method is based on 
complex dilatation techniques (see \cite{kp} for an introductory textbook presentation). The second method is 
based on the Borel resummation of perturbative series, and it has been used mostly to compute resonances of anharmonic oscillators.  
This second method is not always guaranteed to work, since there might be additional contributions to the resonant eigenvalues 
from non-perturbative sectors \cite{alvarez-casares2}. However, it leads to correct results in many cases (like for example for the resonances of 
the cubic oscillator, as proved in \cite{calicetiodd}). The 
results of \cite{gu-s} suggest that, in the case 
of quantum curves, the Borel resummation of the perturbative series leads to the correct spectrum without further ado. We will consider both methods in some detail and, 
as we will see, we will obtain results in perfect agreement with the predictions of topological string theory. One unconventional aspect of our analysis is that we consider 
perturbative expansions around complex points in moduli space, leading to perturbative series with complex coefficients. Many of the complex resonances that we calculate 
are not obtained by lateral Borel resummation of a non Borel-summable real series, as it happens in the conventional cubic and quartic oscillators, 
but by the conventional Borel resummation of a complex Borel-summable series. 

In general, Hamiltonians with complex parameters lead to complex spectra, but this is not always the case. A famous example is the pure 
cubic oscillator with Hamiltonian 
\be
\label{pt-cubic}
\mH= {\mm^2 \over 2} +  \ri \mx^3. 
\ee
Due to PT-symmetry (see e.g. \cite{bender-review}), the spectrum of this Hamiltonian is purely real, as originally conjectured by 
Bessis and Zinn--Justin. In many examples, the reality of the 
spectrum can be already seen in perturbation theory. What is in fact intriguing is the reappearance of complex eigenvalues in some PT-symmetric 
Hamiltonians, due to the spontaneous breaking of PT symmetry. For example, the Hamiltonian 
\be
\label{dt-ham}
\mH= {\mm^2 \over 2} +  \ri \mx^3+\ri \alpha \mx
\ee
has a complex spectrum when $\alpha$ is sufficiently negative. This turns out to be a genuinely non-perturbative effect 
in which small exponential terms become dominant for a certain range of parameters \cite{delabaere-trinh,ben-ber}. 
The theory of quantum curves leads to Hamiltonians with similar properties (see e.g. \cite{koroteev}). In this paper we focus on a particular limit of the 
theory of \cite{ghm,cgm} which was studied in detail in \cite{gm-deformed}. The resulting Hamiltonians have the form 
\be
\mH= \cosh(\mm) + V_N(\mx), 
\ee
where $V_N(x)$ is a polynomial of degree $N$. Appropriate choices of $V_N(\mx)$ lead to PT-symmetric 
Hamiltonians which display similar physical phenomena. One can use the 
conjectural exact quantization conditions (EQCs) predicted in \cite{gm-deformed} to study PT symmetry and its spontaneous breaking in great detail. This study can be regarded as a precision test of the predictions of \cite{gm-deformed}, and as a new family of examples in the world of PT-symmetric models. 

This paper is organized as follows. In section \ref{resqc} we study non-Hermitian operators in the theory of quantum mirror curves. We present two basic 
methods to compute complex spectra in quantum mechanics, we apply them to the operators obtained in the theory of quantum curves, and we compare the results 
to the predictions of the TS/ST correspondence. In section \ref{pt-sec} we consider PT-symmetric operators appearing in the 
theory of quantized SW curves.
In particular, we consider a deformed version of the cubic oscillator studied by Delabaere and Trinh \cite{delabaere-trinh}. 
We show that this oscillator displays the same pattern of PT symmetry breaking, and we show that this is in complete agreement with the exact quantization conditions 
obtained in \cite{gm-deformed} for this system. In section \ref{conc} we conclude with some open problems for the future. Appendix \ref{pf-sec} contains 
some technical details on the WKB analysis of the PT symmetric, deformed cubic oscillator.

\sectiono{Resonances and quantum curves}
\label{resqc}

\subsection{Resonances and non-Hermitian quantum mechanics}

In conventional quantum mechanics, we typically require that operators describing observables are symmetric and self-adjoint. There are however 
important situations in which this is not the case. We might have for example potentials which do not support bound states, but lead to 
{\it resonant states} (also called Gamow vectors) and to a complex spectrum of {\it resonances} (see e.g. \cite{delamg} for a pedagogical 
overview, and \cite{moiseyev} for physical applications).  
More generally, one can consider operators which are not even symmetric (involving for example complex parameters) 
and define in an appropriate way a spectral problem leading to complex eigenvalues. In many interesting examples, resonances are particular 
cases of this more general spectral problems. In this paper, we will 
refer to the complex eigenvalues of a spectral problem as resonances, whether they come from the resonant spectrum of a symmetric Hamiltonian, or from the 
complex spectrum of a non-symmetric Hamiltonian. 

\begin{figure}[tb]
\begin{center}
\begin{tabular}{cc}
\resizebox{60mm}{!}{\includegraphics{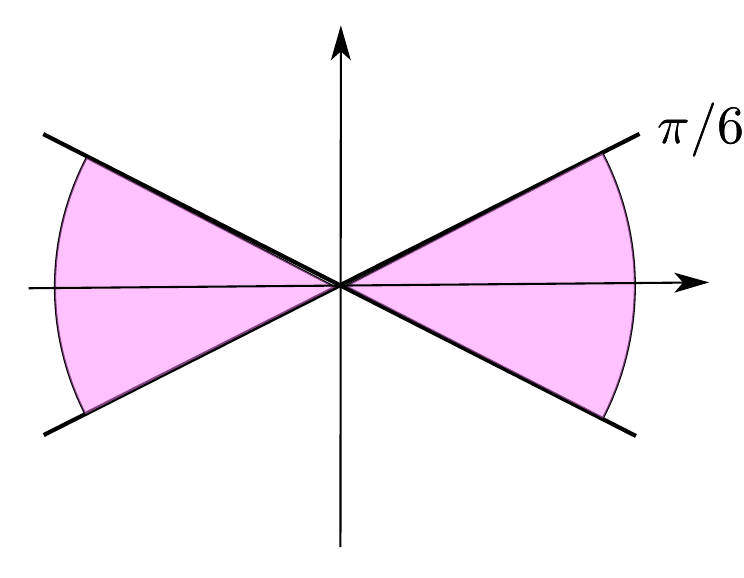}}
\hspace{10mm}
&
\resizebox{60mm}{!}{\includegraphics{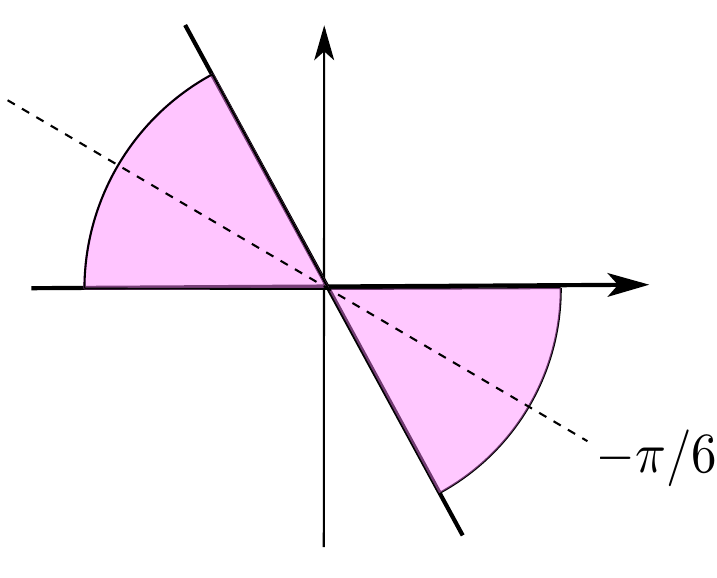}}
\end{tabular}
\end{center}
  \caption{On the left, we depict the regions in the complex $x$ plane where the eigenfunctions of the quartic oscillator with $g>0$ decay at infinity. As $g$ becomes complex, we can define a spectral problem by requiring that the eigenfunctions decay at infinity in the rotated wedges (\ref{rot-wedge}) of the complex plane. The figure on the right shows the relevant wedges for $g<0$. 
}
\label{regions-fig}
\end{figure}

The most canonical example of non-Hermitian quantum mechanics is perhaps the quartic anharmonic oscillator, with Hamiltonian 
\be
\label{qH}
\mH= {\mm^2 \over 2} + {\mx^2 \over 2} + g \mx^4. 
\ee
If $g>0$, this is an essentially self-adjoint operator on $L^2(\IR)$ with a trace-class inverse, therefore $\mH$ has a real, discrete spectrum. 
The behavior at infinity of the eigenfunctions can be obtained with WKB estimates, and one finds 
\be
\psi(x) \approx  \exp \left[ \mp {1 \over 3} {\sqrt{g \over 2}} x^3 \right], \qquad x\rightarrow \pm \infty.
\ee
In fact, the eigenfunctions exponentially decrease in a wedge around the real axis defined by 
\be
\left|{\rm arg}(\pm x)\right|<{\pi \over 6}, 
\ee
 and shown in \figref{regions-fig}. Now suppose that we start rotating $g$ counterclockwise into the complex plane, with an angle $\varphi_g$:
\be
g=|g| \re^{\ri \varphi_g}. 
\ee
Then, the wedges where the WKB wavefunction decreases at infinity rotate clockwise, and are now defined by 
\be
\label{rot-wedge}
\left|{\rm arg}(\pm x)+ {1\over 6} \varphi_g \right|<{\pi \over 6}. 
\ee
The center of these rotated regions is at an angle 
\be
\varphi_x= -{\varphi_g \over 6}. 
\ee
For example, if $\varphi_g= \pi$, we have to rotate the wedge region an angle $-\pi/6$, as shown in \figref{regions-fig}. 
We can now consider the following spectral problem, for arbitrary complex $g$: the eigenfunctions $\psi(x)$ should be solutions of the Schr\"odinger equation for the quartic potential, 
and satisfy
\be
\lim_{|x| \rightarrow \infty} \psi(x)=0
\ee
inside the wedge (\ref{rot-wedge}). This defines a discrete spectrum of eigenvalues $E_n(g)$, $n=0,1,2,\cdots$, 
for any complex $g$ \cite{bw,simon-bw,bender-turbiner}. Note that, when 
$g$ is complex, the Hamiltonian (\ref{qH}) is not even symmetric. When $g$ is real but negative, the Hamiltonian (\ref{qH}) is symmetric but it is not 
essentially self-adjoint (see e.g. \cite{hall-book}). One can however 
define resonant states or Gamow vectors by imposing so-called Gamow--Siegert boundary conditions (see \cite{delamg} for a pedagogical review). 
The spectrum of resonances obtained in this way agrees with the spectrum defined above for general complex $g$, when we specialize it to $\varphi_g=\pi$.

Bender and Wu \cite{bw} 
discovered that the functions $E_n(g)$ obtained in this way display a rich analytic structure: when $|g|$ is large enough, 
they define {\it multivalued} functions of the coupling with three 
sheets (in particular, $E_n(g)$ has $6 \pi$ periodicity as a function of the argument of $g$). As $|g|$ decreases, one finds an infinite set of branch points, 
called {\it Bender--Wu branch points}. 
Asymptotically, the branch points form a two-dimensional quasi-lattice in the complex plane. At each of these branch points, two energy levels coalesce. 
This implies in particular that one can go from one energy level to another by a change of sheet in a single multivalued function. 

A similar structure is found in the cubic oscillator with 
Hamiltonian
\be
\label{cubic-oscillator}
\mH= {\mm^2 \over 2} + {\mx^2 \over 2} - g \mx^3. 
\ee
When $g$ is real, the operator is symmetric but it is not essentially self-adjoint \cite{calicetiodd}. An appropriate definition of the spectral problem for 
arbitrary complex $g$ leads again to a multivalued structure in the energy levels, and to Bender--Wu branch points \cite{alvarezcubicbw} similar to those 
obtained in \cite{bw}. When $g$ is real, the complex eigenvalues obtained in this way agree with the resonances defined by Gamow--Siegert boundary conditions.  

In order to unveil these beautiful structures, it is important to be able to compute the complex eigenvalues $E_n(g)$ from 
first principles in quantum mechanics. We will consider two different computational methods. The first one is the complex dilatation method, which is non-perturbative 
and is implemented numerically by combining it with the Rayleigh--Ritz variational method. The second one is perturbative and is based on the resummation of 
asymptotic series. 

Let us first describe the complex dilatation method (see e.g. \cite{kp} for a general introduction and \cite{bender-resonance} for the original application to the cubic oscillator). 
Let $\mO (\mx, \mm)$ be an operator in one-dimensional quantum mechanics, which is a function of the 
Heisenberg operators $\mx$, $\mm$. One considers the group of complex dilatations
\be
\mU_{ \ri \theta} \mx \mU_{\ri \theta}^{-1} = \re^{\ri \theta}\mx, \qquad \mU_{\ri \theta} \mm \mU^{-1}_{\ri \theta} = \re^{-\ri \theta} \mm. 
\ee
In principle $\theta\in \IC$. Let us now consider the ``rotated" operator, 
\be
\label{Htheta}
\mO_ {\theta}= \mU_{ \ri \theta} \mO  \mU^{-1}_{\ri \theta} =\mO \left( \re^{\ri \theta}\mx,  \re^{-\ri \theta}\mm \right).
\ee
In appropriate circumstances, this operator will have square-integrable eigenfunctions with complex eigenvalues, which are independent of the precise value of 
$\theta$ beyond a given threshold. A numerical approximation to the resulting eigenvalues can be obtained by using the standard Rayleigh--Ritz method. To do this, 
one chooses a basis of $L^2(\IR)$ depending on a given parameter $\omega$, which we will denote as $\{ |\varphi_n (\omega) \rangle \}_{n=0,1, \cdots}$. A convenient 
choice is the basis of eigenfunctions of the harmonic oscillator with normalized mass $m=1$ but depending on the frequency $\omega$. Then, one considers the matrix 
\be
\langle \varphi_n(\omega)| \mO_ {\theta}|\varphi_m (\omega) \rangle, \qquad n,m=0,1,2, \cdots
\ee
One can truncate this matrix at a given level $N$, and calculate its eigenvalues $\lambda_n(N, \theta, \omega)$, $n=0, \cdots, N$. For $\theta$ beyond the threshold, one has
\be
\lim_{N \to \infty} \lambda_n (N, \theta, \omega)= \lambda_n, 
\ee
which are the complex eigenvalues one is looking for. In addition, by changing the value of the angle $\theta$ one can explore the multivalued structure of the energy levels. 
In practice, one calculates the truncated versions $\lambda_n (N, \theta, \omega)$, which 
depend on the size of the matrix, the threshold angle, and the variational 
parameter $\omega$. In many cases precision is dramatically improved, for a given truncation level $N$, by a judicious choice of $\omega$. Different criteria have 
been proposed in the literature to determine 
the optimal value of $\omega$. For example, one criterion is to extremize the trace of the truncated matrix, i.e.\ to choose an $\omega_{\rm opt}$ such that \cite{optimal1, optimal2}
\be
\label{optimal-om}
{\partial \over \partial \omega}\sum_{n=0}^N \langle \varphi_n(\omega)| \mO_ {\theta}|\varphi_n (\omega) \rangle\biggl|_{\omega= \omega_{\rm opt}}=0. 
\ee
This criterion turns out to be very efficient in conventional quantum mechanics. 
In the study of resonances for quantum mirror curves, the stationarity 
condition is more difficult to implement, since there are typically many values of $\omega$ satisfying (\ref{optimal-om}) 
and one has to do a careful study to determine the best choice of variational parameter. 

Another method to compute resonances in quantum mechanics is to perform Borel resummations of perturbative series for energy levels. 
In many cases, the energy of the $n$-th level of a quantum-mechanical system has an asymptotic expansion of the form, 
\be
\label{en-as}
E_n(\lambda)\sim b(\lambda) + \varphi(\lambda), \qquad  \varphi(\lambda)= \sum_{k \ge 0} a_k(n) \lambda^{k}.  
\ee
Here, $\lambda$ is a small control parameter (it could be a coupling constant or the 
Planck constant $\hbar$) and the coefficients 
$a_k(n)$ diverge factorially, for large $k$ and fixed $n$,
\be
|a_k(n)| \sim k!, \qquad k \gg 1. 
\ee
In (\ref{en-as}), $b(\lambda)$ is a $\lambda$-dependent constant which might arise when writing the Hamiltonian as a perturbed 
harmonic oscillator (see the first term in (\ref{eosim}) for an example). One can then define the Borel transform of the formal power series above as 
\be
\widehat \varphi(\zeta)= \sum_{k \ge 0} {a_k(n) \over k!}\zeta^{k}. 
\ee
The Borel resummation of the original power series is then given by 
\be
 \label{analytic}
s(\varphi)(\lambda)=\int_0^{\infty}  \re^{-\zeta} \widehat \varphi (\lambda \zeta) \, \rd \zeta =\lambda^{-1} \int_0^{\infty} \re^{-\zeta/\lambda} \widehat \varphi (\zeta) \, \rd \zeta, 
\ee
where we have assumed that $\lambda>0$ (more general values of $\lambda$ can be addressed by rotating the integration contour appropriately). 
If this integral is well-defined, we say that the original series is Borel summable. A typical obstruction to the existence of this integral is the presence of singularities of the Borel transform 
along the positive real axis. When this is the case, one can still define a generalized Borel resummation along the direction $\theta$, as follows
\be
\label{thetaresiduum}
s_\theta (\varphi)(\lambda) =\int_0^{\re^{\ri \theta} \infty} \re^{- \zeta} \widehat \varphi (\lambda \zeta) \, \rd \zeta. 
\ee
In particular, if $\theta=\pm \delta$, with $0<\delta\ll1$, the integration contour in 
(\ref{thetaresiduum}) can be deformed to a contour $\CC_\pm$ just above or below the 
positive real axis, respectively. The resulting generalized Borel resummations are called {\it lateral Borel resummations} and denoted by 
\be
\label{lateralborel}
s_\pm (\varphi)(\lambda)=\lambda^{-1} \int_{\CC_{\pm}} \rd \zeta \, \re^{-\zeta/\lambda} \widehat \varphi (\zeta).
\ee
In favorable cases, (lateral) Borel resummations of the perturbative series make it possible to recover the exact eigenvalue or resonance $E_n(\lambda)$. 
A well-known example is the cubic oscillator with Hamiltonian (\ref{cubic-oscillator}).
The energy levels have an asymptotic expansion of the form (\ref{en-as}), where the control parameter is $\lambda=g^2$. When $g$ is real, 
the resulting series is not Borel 
summable, but the lateral Borel resummations are well-defined and they give precisely the resonances of the cubic oscillator 
(see e.g. \cite{alvarez-cubic,calicetiodd}). In some cases, however, one has to supplement the perturbative series with a full trans-series including 
exponentially small corrections. This happens for example in the resonances of the cubic oscillator when ${\rm Im}(g)<0$ \cite{alvarez-casares2}. 

In order to build up the perturbative series (\ref{en-as}), one typically has to write the Hamiltonian as a perturbation of a harmonic oscillator. In the case of resonances, this oscillator might have a {\it complex} frequency, and even if the original Hamiltonian has real coefficients, one obtains in this way a perturbative series with complex coefficients. When this series is Borel resummable, one can use conventional Borel resummation to obtain a complex, resonant eigenvalue. 
An elementary example of such a situation is the Hamiltonian 
\be
\mH= {\mm^2 \over 2} - \mx -g^2 {\mx^3 \over 3}, \qquad [\mx, \mm]= \ri \hbar. 
\ee
We can write it as a perturbed harmonic oscillator by introducing the operator
\be
\mz= g^{1/2}\left( \mx -{\ri \over g} \right), 
\ee
so that 
\be
\mH= -{2 \ri \over 3 g} +\mH_2-{g^{1/2} \over 3} \mz^3, \qquad [\mz, \mm]= \ri g^{1/2} \hbar. 
\ee
Here, 
\be
\mH_2={\mm^2 \over 2} -\ri \mz^2 
\ee
is the Hamiltonian of a harmonic oscillator with frequency
\be
\label{cubic-om}
\omega^2_c= -2 \ri. 
\ee
We can now use standard perturbation theory to write down an asymptotic expansion for the energy levels of this Hamiltonian. For the ground state energy we find, by using 
the BenderWu package \cite{bwpack}, 
\be
\label{eosim}
E_0(g)\sim  -{2 \ri \over 3 g} + {1-\ri \over 2} + {11 g \over 288}-{155\over 27648} (1-\ri) g^2+ \cdots, 
\ee
where we set for simplicity $\hbar =g^{-1/2}$. This series is Borel summable along the real axis and it can be easily resummed by using Borel--Pad\'e techniques. One finds for example, with high precision, 
\be
E_0(1)= 0.533167708417867457252...-1.170055989168924629250... \ri, 
\ee
in agreement with a calculation using complex dilatation techniques. Let us note that, when we have a resonance, the sign of the imaginary part of the eigenvalue is not uniquely defined, since it depends on an underlying choice of branch cut. In the case of the resonances of the cubic oscillator (\ref{cubic-oscillator}), this choice corresponds to the different lateral Borel resummations of a non-Borel resummable series. In the example above, what determines the sign is the choice of square root in the frequency of the oscillator (\ref{cubic-om}). 

As emphasized in \cite{gmz}, there is no 
guarantee that the resummation of a Borel-resummable series reproduces 
the exact value that one is after. For example, the perturbative WKB series of the PT-symmetric cubic oscillator (\ref{dt-ham}) is Borel 
resummable, but its resummation misses important non-perturbative effects \cite{delabaere-trinh} (see \cite{ims} for a detailed illustration). 
However, the perturbative series for the energy levels obtained in stationary perturbation theory are often Borel summable to the right eigenvalues. The same phenomenon has been observed 
for quantum mirror curves in the case of conventional bound states with real energies \cite{gu-s}\footnote{As shown in \cite{serone}, even when the perturbative series is not Borel summable, one can often deform the Hamiltonian so as to obtain a Borel summable series which leads to the correct spectrum.}.

\subsection{Quantum mirror curves}

In this section we will focus on the spectral theory of the operators obtained by quantization of mirror curves. 
This theory was first considered in \cite{km,hw}, and a complete conjectural description 
of the exact spectrum of these operators was put forward in \cite{ghm,cgm}. 
The starting point is the mirror curve $\Sigma$ of a toric CY manifold $X$, which can be 
written as a polynomial in exponentiated variables:
\be
\label{mc}
W_X(\re^x, \re^y)=0. 
\ee
This equation can be written in many ways, but in order to extract the relevant operator one should use an appropriate canonical form, as described in \cite{ghm,cgm}. 
The mirror curve depends on the complex structure moduli of the mirror CY. It turns out that these moduli are of two types \cite{hkp,hkrs}: 
$g_{\Sigma}$ ``true" moduli, where $g_\Sigma$ is the genus of the curve, and a certain number of mass parameters $\xi_k$, $k=1, \cdots, r_\Sigma$. 
The quantization of the curve (\ref{mc}) involves promoting $x$, $p$ to Heisenberg operators $\mx$, $\mm$ on $\IR$, with the standard commutation 
relation 
\be
[\mx, \mm]= \ri \hbar. 
\ee
In this way, one obtains an operator $\mO_X$ on $L^2(\IR)$ (when $g_\Sigma>1$, one obtains in fact $g_\Sigma$ different operators, as explained in \cite{cgm}, but for simplicity we will 
focus for the moment being on the case $g_\Sigma=1$). The mass parameters appear as coefficients in the operator $\mO_X$, and its spectral properties 
depend crucially on their values, and in particular on their reality and positivity properties. 
There is a range of values of the parameters in which $\mO_X$ is self-adjoint and its inverse 
\be
\rho_X=\mO_X^{-1}
\ee
 is of trace class. Outside this range, the operator is no longer self-adjoint.

For concreteness, let us consider two operators which will be our main focus in this section. 
The first operator is associated to the toric CY known as local $\IF_0$, and it reads:
\be
\label{localf0}
\mO_{\IF_0}= \re^{\mm} + \re^{-\mm}+ \re^{\mx} + \xi_{\IF_0} \re^{-\mx}. 
\ee
This operator involves a mass parameter $\xi_{\IF_0}$. The operator is self-adjoint and has a trace class inverse as long as 
$\xi_{\IF_0} >0$ \cite{kama}. A closely related example is the operator associated to the local $\IF_2$ geometry, 
\be
\label{localf2}
\mO_{\IF_2}=\re^{\mx} + \re^{\mm}+ \re^{-2 \mx - \mm}+ \xi_{\IF_2}  \re^{-\mx}. 
\ee
This operator is self-adjoint and it has a trace class inverse as long as $\xi_{\IF_2} >-2$ \cite{kmz}. In fact, for this range of parameters, the operator (\ref{localf2}) is equivalent 
to the operator (\ref{localf0}), in the following precise sense \cite{kama,gkmr}. Let $\lambda_{\IF_0}$ be an eigenvalue of the operator 
$\mO_{\IF_0}$ with parameter $\xi_{\IF_0}$. Consider 
now the operator $\mO_{\IF_2}$ with parameter
\be
\label{mass-rel}
\xi_{\IF_2}= \xi_{\IF_0}^{1/2}+ \xi_{\IF_0}^{-1/2}. 
\ee
Then, the eigenvalues $\lambda_{\IF_2}$ of $\mO_{\IF_2}$ are given by
\be
\label{eigen-rel}
\lambda_{\IF_2}= \xi_{\IF_0}^{-1/4} \lambda_{\IF_0}. 
\ee

What happens when the mass parameters are outside the range in which the operator is self-adjoint and has a trace-class inverse? Let us consider for example (\ref{localf0}). 
When $\xi_{\IF_0} <0$, we are in a situation similar to the quartic oscillator with negative coupling, in the sense that the operator is symmetric but does not support bound states. As in conventional 
quantum mechanics, it should be possible to perform an analytic continuation of the eigenvalue problem leading to a {\it discrete} spectrum of resonant states, 
with complex eigenvalues. This should be also the case for more general, complex values of $\xi_{\IF_0}$. We expect to have a similar situation for the operator (\ref{localf2}) 
when $\xi_{\IF_2}<-2$. There are two 
indications 
that this is in fact what happens, as discussed in \cite{cgum}. The first one is that, in some cases, the spectral traces of the inverse operators 
$\rho_{\IF_0, \IF_2}= \mO_{\IF_0, \IF_2}^{-1}$ can be computed in closed 
form as functions of $\xi_{\IF_0}$ or $\xi_{\IF_2}$. For example, one has, for $\xi_{\IF_0}>0$ and $\hbar=2 \pi$, 
\be
\label{first-trace}
\tr \, \rho_{\IF_0}={1\over 8 \pi} {\log(\xi_{\IF_0}) \over \xi_{\IF_0}^{1/2}-1}. 
\ee
This function of $\xi_{\IF_0}$ can be analytically continued to the complex plane, where it displays a branch cut along the negative real axis. The second indication 
is that the TS/ST correspondence predicts a discrete spectrum of complex resonances for the operators $\mO_X$, for any complex value of the mass parameters, as 
explained in \cite{cgum}. This spectrum can be 
computed from the conjectural form of the spectral determinants proposed in \cite{ghm,cgm}. 
In the case of genus one curves, one can also use the EQC proposed in \cite{wzh}. 
Unfortunately, the predictions of the TS/ST correspondence for resonant eigenvalues were not 
verified against first principle calculations in spectral theory. One of the goals of the 
present paper is to fill this gap by developing computational methods to compute resonances, directly in the spectral theory.

Let us briefly review what are the predictions for the spectrum from topological string theory (we refer the reader to the original papers \cite{ghm,cgm,wzh} for more 
details and explanations, and to \cite{mmrev} for a review). The spectrum of the operator $\rho_X$ can be found by looking at the zeros of its spectral determinant, 
\be
\Xi_X (\kappa)={\rm det}\left(1+ \kappa \rho_X\right). 
\ee
When $\rho_X$ is trace class, this is an entire function of $\kappa$ and it has an expansion around $\kappa=0$ given by 
\be
\Xi_X (\kappa)= 1+\sum_{N =0}^\infty Z_X(N) \kappa^N, 
\ee
where $Z_X(N)$ are the {\it fermionic spectral traces} of $\rho_X$ and can be computed from the conventional spectral traces $\tr\, \rho^\ell_X$. 
According to the TS/ST correspondence, these traces can be obtained as Laplace transforms, 
\be
\label{stN}
Z_X(N)= {1\over 2 \pi \ri} \int_{\CC} \re^{\mJ_X(\mu) - \mu N}  \rd \mu.  
\ee
In this equation, $\CC$ is an Airy-like contour of integration, and $\mJ_X(\mu)$ is the so-called grand potential of topological string theory on $X$, which can be explicitly obtained from the BPS invariants of $X$, as explained in \cite{ghm,cgm, mmrev}. 

When the 
mirror curve has genus one, the condition for the vanishing of the spectral determinant can be written as an EQC of the form \cite{wzh}, 
 \be
 \label{eqc}
 {r C t^2 \over2} + D(\boldsymbol{\xi}) t + B(\boldsymbol{\xi}, \hbar)+ \hbar \left( f_{\rm NS}\left( t, \boldsymbol{\xi}, \hbar \right) + f_{\rm NS} \left( {2 \pi t \over \hbar}, 
 \boldsymbol{\xi}^{2 \pi  \over \hbar} ,\frac{4\pi^2}{\hbar}\right) \right)= 2 \pi \hbar \left( n+{1\over 2} \right), \ee
where $n=0, 1, 2, \cdots$ is a non-negative integer. In this equation, $\boldsymbol{\xi}=(\xi_1, \cdots, \xi_{r_\Sigma} )$ is the vector of mass parameters, and 
we denote $\boldsymbol{\xi}^{2 \pi \over \hbar}=(\xi_1^{2 \pi \over \hbar}, \cdots, \xi_{r_\Sigma} ^{2 \pi \over \hbar})$. 
The function $B(\boldsymbol{\xi}, \hbar)$ has the form, 
 \be
 B(\boldsymbol{\xi}, \hbar) = B \left( 1+ {\hbar^2 \over 4 \pi^2} \right)+ b (\boldsymbol{\xi}), 
 \ee
$C$, $r$, and $B$ are constant coefficients depending on the geometry under consideration, $D(\boldsymbol{\xi})$ and $b(\boldsymbol{\xi})$ are functions of the mass parameters, 
and $t$ is related to the true modulus of the curve, 
\be
\kappa=\re^E, 
\ee
 through the so-called quantum mirror map \cite{acdkv}: 
 \be
 t= t(E, \boldsymbol{\xi}, \hbar). 
 \ee
The function $f_{\rm NS}(t,\boldsymbol{\xi}, \hbar)$ can be expressed in terms of the Nekrasov--Shatashvili (NS) limit of the refined topological string free energy \cite{ns}, which we will 
call NS free energy and will denote by $F_{\rm NS}(t, \boldsymbol{\xi}, \hbar)$. The precise relation is, 
\be
f_{\rm NS}(t, \boldsymbol{\xi}, \hbar)= r {\partial F_{\rm NS}^{\rm inst} \over \partial t}.  
\ee
The superscript indicates that we only keep the ``instanton" part of the NS free energy, which can be computed by knowing the BPS invariants of the CY $X$. We also recall that 
\be
F_{\rm NS}(t,\boldsymbol{\xi}, \hbar) ={1\over \hbar} F_0(t,\boldsymbol{\xi}) + \CO(\hbar), 
\ee
where $F_0(t,\boldsymbol{\xi})$ is the genus zero free energy of the toric CY in the large radius frame,
\be
F_0(t,\boldsymbol{\xi}) ={C\over 6} t^3 + D(\boldsymbol{\xi}) t + F_0^{\rm inst}(t,\boldsymbol{\xi}). 
\ee
The equation (\ref{eqc}) determines energy levels $E_n$, $n=0,1,\cdots$, for the modulus $E$, while the eigenvalues of $\rho_X$ are given by $\re^{-E_n}$. 

In the following, it will be useful to understand the operators obtained by quantization of mirror curves from the point of view of 
perturbation theory. To this end, we expand them 
around the minimum $(x_0, p_0)$ of the corresponding function. In this way we will 
make sure that they can be regarded as perturbed harmonic oscillators. By doing this, we shift
\be
\mx \rightarrow \hbar^{1/2} \mx- x_0, \qquad \mm \rightarrow \hbar^{1/2} \mm- p_0, 
\ee
so that the operator is written as 
\be
\mO= s(\xi) \mS, 
\ee
where $s(\xi)$ is a function of the mass parameter, and the operator $\mS$ can be expanded as \cite{gu-s}
\be
\mS=f(\xi) + {\hbar \over 2} \left( \alpha(\xi) \mx^2 + \beta (\xi) \mm^2+ \gamma(\xi) (\mx \mm+ \mm \mx) \right)+ \CO(\hbar^{3/2}), 
\ee
where $f(\xi)$ is a $c$-number. 
One can now perform a linear canonical transformation to eliminate the cross term $\mx \mm+ \mm \mx$, and find in this way a harmonic oscillator with frequency
\be
\omega^2_c= \alpha(\xi) \beta(\xi)-\gamma^2(\xi). 
\ee
The extended version of the BenderWu program due to \cite{gu-s} makes it possible to calculate a formal series expansion in powers of $\hbar$ 
for the eigenvalues of $\mO$:
\be
{1\over s(\xi)} \re^{E_n}\sim f(\xi) + {\hbar \omega_c \over 2} \left(n+{1\over 2} \right)+ \CO(\hbar^2), \qquad n=0,1,\cdots
\ee
In appropriate conditions, the Borel resummation of this series should give the correct value of the (in general complex) eigenvalue $E_n$. 

%
%

\subsection{Resonances in local $\IF_0$}

The TS/ST correspondence of \cite{ghm,cgm} gives predictions for the spectrum of non-Hermitian Hamiltonians in which the mass parameters take complex values. In principle, 
this spectrum can be found directly in operator theory, by first defining an appropriate analytic continuation of the spectral problem, and then computing this spectrum with an appropriate 
approximation technique. In this paper we will assume that the analytic continuation can be performed rigorously, and then 
we will determine the resulting spectrum with both complex dilatation techniques and Borel resummation techniques.

Let us start the discussion with the operator (\ref{localf0}) associated to local $\IF_0$. As explained above, when $\xi_{\IF_0}<0$, and more generally for complex $\xi_{\IF_0}$, we expect to 
find a resonant spectrum. Indeed, if the EQC (\ref{eqc}) holds also in this regime, it leads to complex eigenvalues. In this example, the parameters and functions appearing in the EQC (\ref{eqc}) are given by, 
\be
r=2, \qquad C=1, \qquad D(\xi_{\IF_0})= -\log(\xi_{\IF_0}), \qquad B={2 \pi^2 \over 3}, \qquad b(\xi_{\IF_0})=0. 
\ee
The quantum mirror map has the structure 
\be
t(E, \xi_{\IF_0}, \hbar)= 2 E -2(1+ \xi_{\IF_0}) \re^{-2 E} +\CO\left(\re^{-4 E}\right). 
\ee
Therefore, the EQC involves both $\xi_{\IF_0}$ and $\log(\xi_{\IF_0})$, which appears in the function $D(\xi)$ and in the argument of $f_{\rm NS}$:
\be
\label{logxi}
\xi_{\IF_0}^{2 \pi \over \hbar} = \exp \left( {2\pi \over \hbar} \log(\xi_{\IF_0}) \right).
\ee
When $\xi_{\IF_0}<0$, this leads to an imaginary part in $E_n$. 

Precise predictions about the resonant spectrum can be also found if we take into account the relationship (\ref{mass-rel}), (\ref{eigen-rel}) between $\mO_{\IF_0}$ and $\mO_{\IF_2}$. 
In spectral theory, this relation is derived in \cite{kmz} when both operators are of trace class, but since the same relationship can be derived for the mass parameters of topological string theory \cite{gkmr}, 
we expect it to hold as well for complex values of $\xi_{\IF_0}$, $\xi_{\IF_2}$, once the analytic continuation is properly implemented. One easy consequence of this relation is the following: if 
\be
\label{xi-phase}
\xi_{\IF_0}=\re^{\ri \phi}, \qquad \phi \in \IR, 
\ee
we have 
\be
\xi_{\IF_2} = 2 \cos(\phi), 
\ee
so that $-2\le  \xi_{\IF_2} \le 2$ and the spectrum of $\mO_{\IF_2}$ is real. It follows from (\ref{eigen-rel}) that
\be
\xi^{-1/4}_{\IF_0}\re^{E_n(\xi_{\IF_0})}
\ee
should be real. In other words, for complex values of $\xi_{\IF_0}$ in the unit circle of the complex plane, i.e.\ of the form (\ref{xi-phase}), we expect 
\be
\label{im-part}
{\rm Im}\left(E_n \right)={\phi\over 4}, 
\ee
independently of $n$ and of $\hbar$. 

The appearance of complex resonances is also natural when we consider the operator (\ref{localf0}) from the point of view of perturbation theory. After expanding around the minimum
\be
p_0=0, \qquad x_0= {1\over 2} \log \xi_{\IF_0}, 
\ee
we find the operator
\be
\mO_{\IF_0}= \re^{\mm} + \re^{-\mm}+ \xi_{\IF_0}^{1/2} \left( \re^{\mx} +  \re^{-\mx}\right). 
\ee
As already pointed out in \cite{hw}, this is a perturbed harmonic oscillator of frequency 
\be
\label{ren-freq}
\omega_c^2= 4 \varpi ^2= 4 \xi_{\IF_0}^{1/2}, 
\ee
where we have introduced the renormalized frequency $\varpi$. When $\xi_{\IF_0}<0$ this oscillator has a complex frequency, 
just as in the example (\ref{cubic-om}). By expanding around this oscillator, one can write down perturbative series for the 
energy levels. For example, one finds, for the ground state energy, 
\be
\label{pert-series}
\re^{E_0} = \left(2 \varpi ^2+2\right)+\varpi  \hbar +\frac{1}{16} \left(\varpi
^2+1\right) \hbar ^2-\frac{3 \varpi ^4-10 \varpi ^2+ 3
	}{768 \varpi }\hbar ^3+\CO(\hbar^4). 
	\ee
The structure (\ref{im-part}) for mass parameters in the unit circle can be also deduced from the explicit form of perturbation theory, since in the case (\ref{xi-phase}) the renormalized frequency 
$\varpi$ introduced in (\ref{ren-freq}) is also a pure phase, $\varpi= \re^{\ri \phi/4}$, and the perturbative series is of the form $\varpi$, times a formally real series (this can be 
easily checked in the first terms of (\ref{pert-series})). If this real series is Borel summable, as it is the case, the only imaginary contribution to its 
Borel--Pad\'e resummation comes from the overall factor $\varpi$. This leads to the imaginary part (\ref{im-part}). 

Let us now present some detailed calculations of the eigenvalues for different values of the parameter $\xi_{\IF_0}$ in the resonant region. We have 
focused on three values of $\hbar$, namely, $2 \pi$, $\pi$ and $2 \pi/3$, where the EQC can be written in a simpler form (the case $\hbar=2 \pi$ is the ``self-dual" or 
``maximally supersymmetric case", where the theory simplifies considerably, as already noted in \cite{ghm}; the simplification in the cases of $\hbar=\pi, 2\pi/3$ follows 
from the results of \cite{szabolcs}). The first computational method is complex 
dilatation, where we use the Rayleigh--Ritz method with the harmonic oscillator basis and an optimal 
value of the frequency satisfying (\ref{optimal-om}). The second method involves the Borel--Pad\'e resummation of the perturbative series, and finally we obtain the 
prediction for spectrum based on the EQC and the TS/ST correspondence. We will focus on the ground state, although we have studied as well excited states.

\begin{table}[tb]
\begin{center}
  \begin{tabular}{|c|l|l|l|}\hline
  	$\hbar$ & \multicolumn{1}{|c}{$2 \pi$}  &  \multicolumn{1}{|c}{$\pi$} & \multicolumn{1}{|c|}{$2 \pi/3$} \\ \hline
	 c.d.   &   $2.7606757803$  &              $1.97575795104945$ & $1.69010343823507799$ \\
	 BP   &    $ 2.7606757803195866$  & $1.97575795104945438307691$ &$1.690103438235077991455$ \\
	EQC   &  $2.7606757803195866 $ &  $1.97575795104945438307691$ & $1.690103438235077991$ \\ \hline
  \end{tabular}
  \vspace{3mm}
  \caption{Real part of the first resonance for local $\IF_0$ with $\log \xi_{\IF_0}= \pi \ri$ and three different values of $\hbar$. Here, c.d. denotes the complex dilatation method, while BP denotes 
  the Borel--Pad\'e method. }
  \label{tab:f0minusone}
\end{center}
\end{table}

Let us first consider values with $|\xi_{\IF_0}|=1$, where we expect the imaginary part of the energy to be given by the simple expression (\ref{im-part}). A simple case is 
$\xi_{\IF_0}=-1$, with the choice of logarithmic branch given by $\log \xi_{\IF_0}= \pi \ri$. The imaginary part should be in this case
\be
{\rm Im}(E_n)={\pi \over 4}. 
\ee
We find that the three methods reproduce this imaginary part with arbitrary precision. The results for the real part of the first resonant state are shown in Table \ref{tab:f0minusone}. In the complex dilatation 
method, we used matrices of rank $200$ for the Rayleigh--Ritz approximation, and in the Borel--Pad\'e resummation we used $150$ terms generated with the BenderWu program. These two methods 
become more efficient for smaller values of $\hbar$. In the EQC, we have used $100$ terms in the instanton expansion. As we see, we obtain perfect agreement between the three methods (in the case 
of $\hbar=2 \pi/3$, the Borel--Pad\'e method and the EQC can be seen to agree up to $35$ digits). Similar results can be found when $\xi_{\IF_0}=\ri$, and with the choice of logarithmic branch 
$\log \xi_{\IF_0}= \pi \ri/2$. The imaginary part should be in this case
\be
{\rm Im}(E_n)={\pi \over 8}. 
\ee
We show some of the results in Table \ref{tab:f0i}. Again, the imaginary part was reproduced with arbitrary precision. In the complex dilatation 
method, we used matrices of rank $200-400$ for the Rayleigh--Ritz approximation (depending on the value of $\hbar$), and in the Borel--Pad\'e resummation we used $100$ terms generated 
with the BenderWu program. In the EQC, we used again $100$ terms in the instanton expansion.

\begin{table}[tb]
\begin{center}
  \begin{tabular}{|c|l|l|l|}\hline
  	$\hbar$ & \multicolumn{1}{|c}{$2 \pi$}  &  \multicolumn{1}{|c}{$\pi$} & \multicolumn{1}{|c|}{$2 \pi/3$} \\ \hline
	 c.d.   &   $2.851995310270838$  &  $2.111510733777354897475$  &  $1.8523966583371164256660$\\
	 BP   &    $2.8519953102708$  & $2.111510733777354897475$ & $1.852396658337116425665972544$\\
	EQC   &  $2.851995310270838$ &  $2.111510733777354897475$  & $1.852396658337116425665972544$  \\ \hline
  \end{tabular}
  \vspace{3mm}
  \caption{Real part of the first resonance for local $\IF_0$ with $\log \xi_{\IF_0}= \pi \ri/2$ and three different values of $\hbar$. }
  \label{tab:f0i}
\end{center}
\end{table}

We have performed tests for values of $\xi_{\IF_0}$ which are not of the form (\ref{xi-phase}), and therefore the imaginary part does not have 
a closed form. For example, in Table \ref{tab:f0minus4} we list the first resonance when $\xi_{\IF_0}=-4$ and $\hbar=\pi$. All methods agree up to the numerical precision 
we have achieved. All these results support the conclusion that the conjecture of \cite{ghm} (or its reformulation in \cite{wzh}) is still valid in the complex realm, when one considers resonances. They also 
support the conjecture that Borel resummation of the perturbative series in $\hbar$ is sufficient to reconstruct the exact answer, without the need of adding additional 
instanton-like corrections in $\hbar$. These corrections are expected to be of order $\exp(-4 \pi^2/\hbar)$ \cite{ghm}, 
i.e. of order $10^{-3}$ for $\hbar=2 \pi$ and of order $10^{-9}$ for $\hbar=2 \pi/3$. Our calculations are precise enough to detect such corrections, if present. In view of our results, instanton corrections 
of this type are very likely to be absent.

\begin{table}[tb]
\begin{center}
  \begin{tabular}{|c|lcl|}\hline
  	$\hbar$  &  \multicolumn{3}{|c|}{$\pi$}    \\ \hline
  	c.d.   &   $2.36465369971376$&$+$&$0.94715399106287~\ri $\\
  	BP   &    $2.364653699713759905588127 $&$+$&$  0.947153991062872399610128~\ri$ \\
  	EQC   &  $2.36465369971375990558812739733 $&$+$&$  0.947153991062872399610128166~\ri$   \\ \hline
  \end{tabular}
  \vspace{3mm}
  \caption{First resonance for local $\IF_0$ with $\log \xi_{\IF_0}= 2 \log(2) + \pi \ri$ and $\hbar=\pi$. }
  \label{tab:f0minus4}
\end{center}
\end{table}

The operator corresponding to local $\IF_2$, given in (\ref{localf2}), can be analyzed very similarly to local $\IF_0$. In this case, the extremum of the 
underlying function $O_{\IF_2}(x, p)$ occurs at 
\be
\label{minF2}
x_0=-p_0= {1\over 2} \log\left(\xi_{\IF_2}+2\right),  
\ee
and when we expand around this value we obtain
\be
\mO_{\IF_2}={\sqrt{\xi_{\IF_2}+2}} \, \mS_{\IF_2}, 
\ee
where
\be
\mS_{\IF_2}= \re^{\mx} +{1\over \xi_{\IF_2} +2} \re^{\mm} + {1\over \xi_{\IF_2}+2} \re^{-2\mx-\mm}+{\xi_{\IF_2}  \over \xi_{\IF_2}+2} \re^{-\mx}. 
\ee
This operator is a perturbed harmonic oscillator with frequency 
\be
\omega_c^2= {4 \over2+\xi_{\IF_2}}. 
\ee
The lack of trace class property for $\xi_{\IF_2}<-2$ is obvious from this description: the minimum (\ref{minF2}) moves to the complex plane, and the frequency of the 
oscillator becomes imaginary. We have studied the spectrum of resonant states for various values of $\xi_{\IF_2}$. Numerically, this operator leads to less precision than (\ref{localf0}), 
but we have obtained again full agreement between the different methods (complex dilatation, Borel resummation and exact quantization conditions).

\subsection{Resonances in perturbed local $\IP^2$}

In \cite{cgum} the following operator was considered, 
\be
\label{pert-localp2}
\mO_{1,1,\xi}= \re^{\mx} + \re^{\mm}+ \re^{-\mx-\mm}+ \xi \re^{-2\mm}. 
\ee
It can be regarded as a perturbation of the operator associated to local $\IP^2$, by the term $\xi \re^{-2 \mm}$. It 
arises by quantization of the mirror curve of the local $Y^{3,0}$ geometry. This curve has genus two, so in order to obtain the spectrum of $\mO_{1,1, \xi}$ one has to consider the higher genus version 
of the TS/ST correspondence presented in \cite{cgm} and further studied in \cite{cgum}. In particular, one has to calculate the relevant spectral determinant, and the quantization condition can not be 
put in the simple form presented in \cite{wzh}. The operator (\ref{pert-localp2}) is trace class for $\xi\ge 0$, and as explained in \cite{cgum} it is expected to display resonances when $\xi<0$. 

Some insight on the structure of this operator can be obtained by writing it as a perturbed harmonic oscillator. In this case, the minimum of the underlying function 
$O_{1,1,\xi}(x,p)$ occurs at 
\be
\re^{x_0}=X_0= {1\over 2^{1/3}} \left( 1+ 4 \xi + {\sqrt{1+ 8 \xi}} \right)^{1/3}, \qquad \re^{p_0}=P_0= {X_0 \over 4} { {\sqrt{1+ 8 \xi}} -1\over \xi}. 
\ee
If we shift $\mx$, $\mm$ around this minimum, we find that
\be
\mO_{1,1, \xi}= X_0 \mS_{\xi}, 
\ee
where
\be
\ba
\mS_{\xi}& =\re^\mx+  \frac{\sqrt{8 \xi +1}-1 }{4 \xi } \re^{\mm} +\frac{8 \xi }{\left(\sqrt{8 \xi +1}-1\right) \left(4 \xi +\sqrt{8 \xi +1}+1\right) }\re^{-\mx -\mm}\\
& +\frac{2 \xi }{\left(4 \xi
   +\sqrt{8 \xi +1}+1\right) } \re^{-2\mx}. 
   \ea
   \ee
   After expanding around $\mx=\mm=0$, we find a harmonic oscillator whose frequency is given by 
\be
\omega_c^2(\xi)= \frac{24 \xi  \left(8 \xi +\sqrt{8 \xi +1}+1\right)}{\left(\sqrt{8 \xi +1}-1\right) \left(4 \xi +\sqrt{8 \xi
   +1}+1\right)^2}.  
   \ee
When
\be
-{1\over 8} < \xi<0
\ee
we have a perturbed harmonic oscillator with a positive frequency, and the perturbative series for the energy levels has real coefficients. 
Since the operator does not have bound states but resonances, we expect that the perturbative series is {\it not} Borel summable, as in the conventional 
cubic oscillator. For example, when $\xi=-1/9$, one finds for the 
perturbative series for the ground state, 
\be
\re^{E_0}= \left( {2\over 3} \right)^{2/3} \left( {15 \over 4} +{3\over 4} \hbar-{79 \over 48} \hbar^2-{12659 \over 576} \hbar^3- {32066045 \over 62208} \hbar^4-\cdots \right), 
\ee
which has non-alternating signs and is not expected to be Borel summable. A study of the Borel plane with Pad\'e approximation techniques indicates indeed the existence of 
a branch cut singularity along the real axis. We expect that resonances are obtained in this case by lateral resummation of a non-Borel summable series. In contrast, when 
\be
\xi < -{1\over 8}
\ee
the perturbative series has complex coefficients, as in the example of local $\IF_0$ with $\xi_{\IF_0}<0$. 

In order to test the TS/ST conjecture for the resonant regime of the operator (\ref{pert-localp2}), we have calculated the spectral traces from the Airy integral formula (\ref{stN}), used these to 
calculate an approximation to the spectral determinant, and finally determined the zeroes of this determinant to locate the resonances. In our calculations we set $\hbar=2 \pi$, we calculated the traces up to $N=10$ and we used an approximation up to seventh instanton order for the grand potential $\mJ_X$ in (\ref{stN}). The results for the first resonance, for $\xi=-1/9$ and $\xi=-9/8$, are presented in Table \ref{table:defp2}, and compared to results obtained by complex dilatation with a matrix of rank $800$ (only stable digits are shown, as usual). Although the precision is smaller than the one achieved in the simpler cases of genus one, we still find full agreement between the predictions of the TS/ST correspondence and a first-principles computation of the resonant spectrum. The result for $\xi=-9/8$ can be also reproduced by using the Borel--Pad\'e resummation of the perturbative series. In the case of $\xi=-1/9$ we have to use lateral 
resummation and the standard Borel--Pad\'e method does not converge well, due to the high value of $\hbar$. It would be interesting to improve on this to obtain these resonances by resummation techniques.

\begin{table}[tb]
\begin{center}
  	\begin{tabular}{|c|lcl|lcl|}\hline
  	$\xi$ & \multicolumn{3}{|c|}{$-1/9$}   &  \multicolumn{3}{|c|}{$-9/8$} \\ \hline
  	c.d.   &   $17.6475073  $&$+$&$  4.7738402~\ri$ & $ 21.4121476$&$+$&$8.3088781~\ri$  \\
  	EQC   &  $17.647507$&$+$&$4.77384~\ri $& $21.41215$&$+$&$\, 8.30888~\ri$    \\ \hline
  \end{tabular}
  \vspace{3mm}
  \caption{Value of $\re^{E_0}$ for the operator (\ref{pert-localp2}) and $\hbar=2 \pi$, for $\xi=-1/9$ (left) and $\xi=-9/8$ (right). }
  \label{table:defp2}
\end{center}
\end{table}

\subsection{Multivalued structure}

In conventional quantum mechanics with polynomial potentials, like (\ref{qH}) and (\ref{cubic-oscillator}), 
the spectrum has a non-trivial multivalued dependence on the complex values of the couplings. For example, as we mentioned in section \ref{resqc}, in the case of the 
quartic oscillator, the energy levels have $6 \pi$ periodicity as functions of the argument of $g$, when $|g|$ is sufficiently large.  

In the case of the operators obtained from mirror curves,  the energy levels depend generically on the logarithm of the mass parameters. Therefore, they should have an infinitely sheeted 
structure as we consider complex values of these parameters, corresponding to the different choices of the branches of the logarithm. 
In the case of local $\IF_0$, for example, this is strongly suggested by exact trace formulae like (\ref{first-trace}), by the structure of the EQC (see e.g. the comment around eq. (\ref{logxi})),  
and also by the equivalence (\ref{mass-rel}) between local $\IF_0$ and $\IF_2$. Indeed, in order to obtain values of $\xi_{\IF_2}$ 
in the resonant region, i.e.\ $\xi_{\IF_2}<-2$, one should consider values of $\xi_{\IF_0}$ of the form 
\be
\log \xi_{\IF_0}= 2 \cosh^{-1}\left( {|\xi_{\IF_2}| \over 2} \right)+ 2 k \pi \ri, 
\ee
where $k$ is an odd integer. This multivalued structure can be also partially seen in the perturbative calculation of the energy levels. If we consider again the example of local $\IF_0$, we see that the determination of the frequency $\omega_c$ requires a choice of branch for $\xi_{\IF_0}^{1/4}$. 

In our calculations in the previous sections we have chosen implicitly the principal branch of the logarithm of the mass parameter, but we can explore other branches. 
This is in principle straightforward in the EQC, where one simply considers different branches of the logarithm. In the complex dilatation method, reaching other sheets 
requires rotating $\theta$ or of the variational parameter $\omega$ in the complex plane. In the Borel resummation technique, one might have access to other sheets by a choice of 
different branches in $\omega_c$ (but only to finitely many). We have performed a preliminary investigation of this multi-sheeted structure. For example, in the case of local $\IF_0$ with $\hbar=2 \pi$, 
one can consider the value $\log(\xi_{\IF_0})= 3 \pi \ri$, instead of $\pi \ri$, as we did in Table \ref{tab:f0minusone}. The EQC gives, 
\be
E_0=1.510433421361...+ {3 \pi \ri \over 4}. 
\ee
It is possible to access this sheet with the complex dilatation method, by changing the value of $\omega$ (this is equivalent to choosing a different value of $\theta$), and one finds:
\be
E_0=1.510433421...+ {3 \pi \ri \over 4}. 
\ee
In the perturbative calculation (\ref{pert-series}), this sheet corresponds to the choice $\varpi = 3 \pi \ri/4$, instead of $\varpi=\pi \ri/4$. Although the usual Borel--Pad\'e method displays in this case an 
oscillatory behavior, the result of the resummation is compatible with the values shown above. 

An additional difficulty to explore the multivalued structure is that, in some cases, the EQC  (\ref{eqc}) does {\it not} converge when one tries to access other sheets for the log of the mass parameter. The EQC 
is given by a power series expansion, which in the trace class case converges for values of the modulus corresponding to the physical spectrum. 
However, in the complex case, the convergence depends on the choice of sheet, of level $n$, and of $\hbar$. 
For example, for $\log(\xi_{\IF_0})=3 \pi \ri$ and $\hbar=\pi$, we have not been able to obtain the first resonance ($n=0$) with the EQC. This raises interesting issues that we discuss in the concluding section 
of the paper.

\sectiono{PT-symmetric quantum curves}
\label{pt-sec}

\subsection{PT symmetry and non-perturbative effects} 

A particularly interesting subset of non-Hermitian 
operators are those which display PT symmetry. A PT transformation changes the Heisenberg operators as 
\be
\mx \rightarrow - \mx, \qquad \mm \rightarrow -\mm, 
\ee
and it also changes the imaginary unit $\ri$ into $-\ri$. Operators which are invariant under this transformation are called PT-symmetric operators. 
A simple example is the Hamiltonian (\ref{pt-cubic}), with a purely cubic, imaginary potential, but more general examples are possible: the cubic 
oscillator (\ref{cubic-oscillator}) leads 
to a PT-symmetric operator when ${\rm Re}(g)=0$. 

When an operator is PT-symmetric, its eigenvalues are either real or come in complex conjugate pairs. In the first case (which is realized in (\ref{pt-cubic})) we say that PT is unbroken, while in the second 
case we say that PT symmetry is broken. It turns out that, in many quantum mechanical models, the breaking of PT symmetry is a non-perturbative effect, 
due to complex instantons. A beautiful example which illustrates the phenomenon is the Hamiltonian (\ref{dt-ham}), studied in detail by Delabaere and Trinh in 
\cite{delabaere-trinh} (a similar model is studied in \cite{ben-ber}, and related considerations on the importance 
of non-perturbative effects in PT symmetry breaking are made in \cite{dt-beyond}). When $\alpha>0$, the eigenvalues of this Hamiltonian are real. However, as $\alpha$ decreases and takes negative values, we find a decreasing sequence of values $\alpha_n<0$, $n=1,2, \cdots$, for which the energy levels $E_{n-1}$ and $E_{n}$ coalesce and become complex conjugates. Therefore, if 
\be
\alpha_{n+1}< \alpha< \alpha_n, 
\ee
the first $2n$ energy levels come in complex conjugate pairs. This phenomenon can be understood quantitatively by using EQCs and 
the complex WKB method \cite{delabaere-trinh} (see also \cite{unfolding} for a closely related analysis of Bender--Wu branch points with exact EQCs). Let us first note that the potential 
\be
\label{pot-alpha}
V(x)= \ri x^3 +\ri \alpha x, 
\ee
has three turning points when $\alpha\ge 0$: one of them, $x_0$, is in the imaginary axis, while the other two, denoted by $x_\pm$, are 
in the fourth and the third quadrant of the complex plane, respectively. Let $\gamma$ be a cycle encircling the turning points $x_0$ and $x_+$. By using the all-orders WKB method, 
one obtains a WKB period associated to $\gamma$, which can be decomposed into real and imaginary parts as 
\be
 \Pi_{\gamma}= {1\over 2}\Pi_{\rm p} - {\ri \over2}  \Pi_{\rm np}, 
 \ee
 where ${\rm p}$, ${\rm np}$ stand for ``perturbative" and ``non-perturbative." The periods $\Pi_{\rm p}$, $\Pi_{\rm np}$ are real formal 
 power series in $\hbar^2$, 
 \be
\Pi_{ {\rm p}, {\rm np}} = \sum_{n \ge 0} \Pi_{ {\rm p}, {\rm np}}^{(n)} \hbar^{2n}, 
\ee
which turn out to be Borel summable \cite{ddpham, dpham, delabaere-trinh}. We will denote by $s(\Pi_{\rm p})$, $s(\Pi_{\rm np})$ their (real) 
 Borel resummations. The Bohr--Sommerfeld (BS) approximation to the spectrum is given by (see e.g. \cite{bb-pt, bender-review})
 \be
 \label{bs}
 \Pi_{\rm p}^{(0)}= 2 \pi \hbar \left(n+{1\over 2} \right), \qquad n=0,1,2, \cdots
 \ee
 This can be promoted to an all-orders, perturbative WKB quantization condition
  \be
  \label{pt-allwkb}
 s\left( \Pi_{\rm p} \right)= 2 \pi \hbar \left(n+{1\over 2} \right), \qquad n=0,1,2, \cdots
 \ee
 The perturbative quantization conditions (\ref{bs}), (\ref{pt-allwkb}) lead to {\it real} spectra, so they can not explain the breaking of PT symmetry observed in this model. 
However, a careful analysis based on the exact WKB method shows that these conditions miss non-perturbative effects due to complex instantons \cite{bpv}. The correct quantization condition is given by 
 \be
 \label{eqc-pt}
 2 \cos\left( {1\over 2 \hbar}  s(\Pi_{\rm p}) \right)+ \re^{- {1\over 2 \hbar} s(\Pi_{\rm np})}=0. 
 \ee
The second term in the l.h.s.~gives a non-perturbative correction to (\ref{pt-allwkb}). This correction is crucial to obtain the right results, but it is typically exponentially small (see \cite{ims} for a recent implementation 
of (\ref{eqc-pt}) in this typical regime). However, as shown in 
\cite{delabaere-trinh, ben-ber}, as $\alpha$ becomes more and more negative, the non-perturbative correction becomes of order one, and the quantization condition (\ref{eqc-pt}) does not 
have solutions for real $E$. Therefore, PT symmetry is broken. 

\begin{figure}[tb]
\begin{center}
\begin{tabular}{cc}
\resizebox{60mm}{!}{\includegraphics{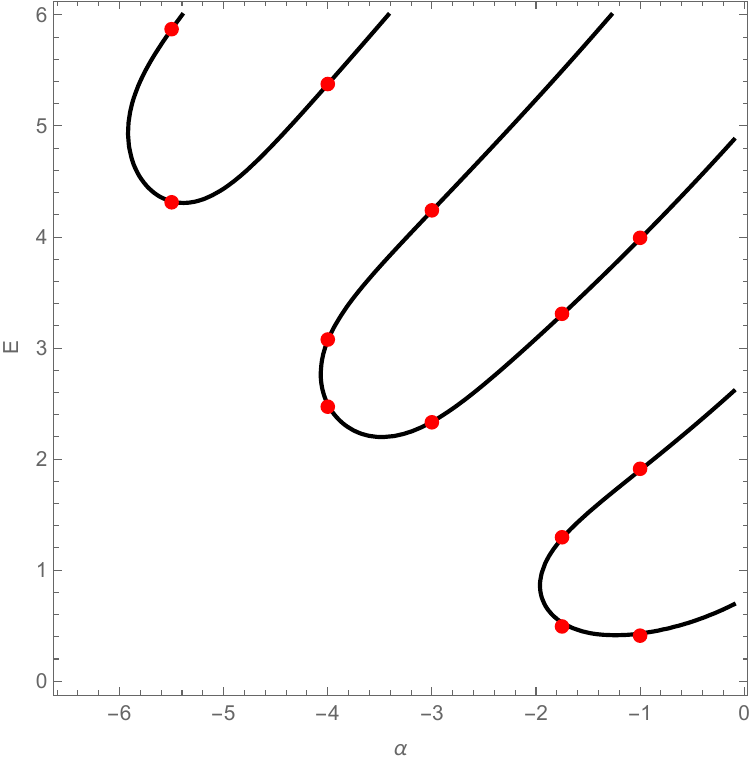}}
\hspace{10mm}
&
\resizebox{60mm}{!}{\includegraphics{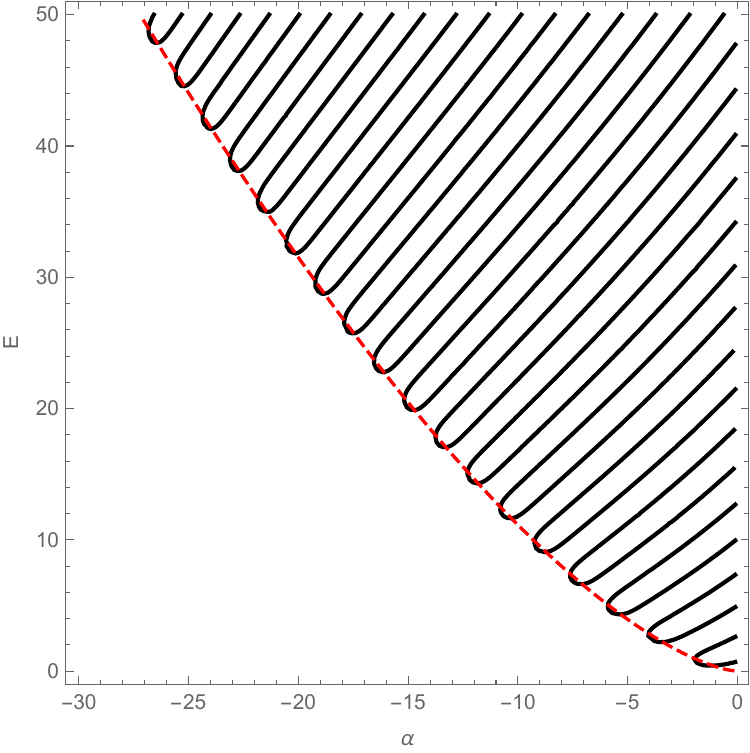}}
\end{tabular}
\end{center}
  \caption{(Left) The curves in the $(\alpha, E)$ plane defined by the approximate quantization condition (\ref{approx-qc}), for $\hbar=1$. The red dots are numerical 
  values for the spectrum of (\ref{dt-ham}) obtained with complex dilatation. (Right) There are clearly two different phases in the $(\alpha, E)$ plane separated by the 
  line (\ref{vanishing}), which is shown here in dashed red. 
}
\label{pt-phases-figs}
\end{figure}

The qualitative structure of PT symmetry breaking is already captured by considering the leading approximation to the full quantization condition (\ref{eqc-pt}), but including the contribution due to 
$\Pi_{\rm np}$ also at leading order, as pointed out in \cite{ben-ber}. This leads to the approximate 
quantization condition 
\be
\label{approx-qc}
 2 \cos\left( {1\over 2 \hbar} \Pi_{\rm p}^{(0)} \right)+ \re^{- {1\over 2 \hbar} \Pi^{(0)}_{\rm np}}=0. 
 \ee
 The functions $ \Pi_{ {\rm p}, {\rm np}}^{(0)}$ can be explicitly written down in terms of elliptic integrals, by using e.g. the results of \cite{cm-ha,ims}. The equation (\ref{approx-qc}) 
 describes the curves of possible energy levels in the $(\alpha, E)$ plane, in very good agreement with numerical results, as shown in \figref{pt-phases-figs} for $\hbar=1$. It is clear that 
 the breaking of PT symmetry leads to two different phases in the $(\alpha, E)$ plane: one phase of real values, and another of complex values. Since the phase transition is triggered by a non-perturbative 
 effect becoming of order one, the boundary separating the two phases can be described approximately by 
 \be
 \label{vanishing}
 \Pi^{(0)}_{\rm np}(\alpha, E)=0, 
 \ee
 as we also illustrate in \figref{pt-phases-figs}. This is reminiscent of large $N$ phase transitions triggered by instantons \cite{neuberger} (see \cite{mmbook} for a textbook exposition), where the 
 transition point is defined by the vanishing of the instanton action as a function of the moduli. A careful analysis of (\ref{approx-qc}) leads for example to a semiclassical, asymptotic 
 formula for the values of $\alpha_n$: 
 \be
 \alpha_n= -x_n^{4/5}, \qquad x_n\sim  (4n-1){\pi \over \omega_0}, \qquad n=1,2, \cdots
 \ee
 where
 \be
 \omega_0 =  \Pi^{(0)}_{\rm p}(1, E_0), 
 \ee
 and the energy $E_0$ is defined as the zero of (\ref{vanishing}) when $\alpha=1$. The sequence of values $\alpha_n$ is a one-dimensional ray of the two-dimensional 
 quasi-lattice of Bender--Wu singularities of the cubic oscillator \cite{alvarezcubicbw}.

\subsection{PT symmetry in deformed quantum mechanics}

In \cite{gm-deformed}, a deformed version of one-dimensional quantum mechanics was introduced. The Hamiltonian of this theory is given by 
\be
\label{q-ham}
\mH_N=\Lambda^N\left(  \re^\mm+\re^{-\mm} \right) +V_N(\mx), 
\ee
where 
\be
V_N(\mx)= \sum_{k=0}^{N-1} (-1)^k \mx^{N-k} h_k 
\ee
is a degree $N$ potential, and we take $h_0=1$, $h_1=0$ without loss of generality. We will refer to $\Lambda$, $h_k$, $k=2, \cdots, N-1$ as 
the moduli of the potential. As explained in detail in \cite{gm-deformed}, the Hamiltonian (\ref{q-ham}) can be obtained by quantizing 
the Seiberg--Witten (SW) curve of $SU(N)$, $\CN=2$ Yang--Mills theory \cite{sw,af,klty}, and in this case the parameters $\Lambda$, $h_k$ correspond to the dynamically generated 
scale and the Coulomb branch moduli, respectively. One interesting aspect of the Hamiltonian (\ref{q-ham}) is 
that its spectrum of resonances or bound states is encoded in an EQC which can be written in closed form for any $N$. As in the conventional 
exact WKB method, this condition can be written in terms of resummed WKB periods of the underlying SW curve:
\be
\label{sw-curve}
\Lambda^N\left(  \re^p+\re^{-p} \right)+  \sum_{k=0}^{N} (-1)^k x^{N-k} h_k=0, 
\ee
where $(-1)^{N-1} h_N$ can be regarded as the eigenvalue of the Hamiltonian (\ref{q-ham}). 
As shown in \cite{mirmor}, these WKB periods are nothing but the quantum periods 
appearing in the NS limit \cite{ns} of $\CN=2$ super Yang--Mills theory. They are denoted by
\be
\label{sw-pers}
\Pi_{A_i}=2 \pi a_i, \qquad \Pi_{B_i}={\partial F_{\rm NS} \over \partial a_i}, \qquad i=1, \cdots, N-1, 
\ee
and they have a formal power series expansion in $\hbar^2$. The zero-th order terms of this expansion are given by the usual SW periods. 
In this equation, $F_{\rm NS}$ is the 
free energy of SW theory in the NS limit of the so-called Omega-background. It has the structure, 
\be
\label{pinst}
F_{\rm NS}=  F_{\rm NS}^{\rm pert} +F_{\rm NS}^{\rm inst}, 
\ee
where $F_{\rm NS}^{\rm pert}$ is the so-called perturbative 
piece, and $F_{\rm NS}^{\rm inst}$ is the instanton piece, which can be computed by using instanton calculus 
in the gauge theory \cite{n}. This leads to an expansion in powers of $\Lambda^{2N}$, of the form 
\be
\label{gauge-exp}
F^{\rm inst}_{\rm NS}= \sum_{\ell \ge 1} F^{(\ell)}_{\rm NS}(\hbar, a_i) \Lambda^{2N \ell}. 
\ee
 In addition to this series, one can also obtain gauge theory 
expansions for the moduli $h_k$ of the SW curve in terms of the A-periods $a_i$, 
\be
\label{hk-gauge}
h_k= h_k^{\rm pert}+ \sum_{\ell \ge 1} h_k^{(\ell)}(\hbar, a_i) \Lambda^{2N \ell},  
\ee
where $h_k^{\rm pert}$ is a classical or perturbative piece which can be determined by elementary algebraic considerations. These 
expansions for the moduli are sometimes called ``quantum mirror maps" and they provide a dictionary between the moduli of the curve 
and the quantum A-periods $a_i$. 
More details, explicit expressions and examples for the series (\ref{gauge-exp}) and (\ref{hk-gauge}) can be found in \cite{gm-deformed}. 
Let us note that, in contrast to the standard WKB expansions in $\hbar^2$, the instanton expansions (\ref{gauge-exp}) and (\ref{hk-gauge}) 
are expected to be convergent in the 
so-called large radius region or semiclassical region of the SW moduli space, 
where the moduli $a_i$ are large in absolute value. They provide explicit resummations of the quantum periods and make unnecessary the 
use of Borel techniques, as emphasized in \cite{gm-deformed}.  

It is clear that, by an appropriate choice of the moduli $\Lambda$, $h_k$, $k=2, \cdots, N-1$, we can easily engineer Hamiltonians of the form (\ref{q-ham}) which are 
PT symmetric. In fact, any PT symmetric Hamiltonian of the form 
\be
\mH= {\mm^2 \over 2} + V(\mx),
\ee
where $V(x)$ is a polynomial potential, leads to a deformed Hamiltonian of the form (\ref{q-ham}). For concreteness, we will focus on the deformed version of the Hamiltonian (\ref{dt-ham}), 
which is obtained by taking $N=3$ and $\Lambda=\ri$. One has 
\be
\label{def-pt}
\mH = \ri \mH_3= 2\cosh(\mm)+ \ri \mx^3+ \ri \alpha \mx, 
\ee
where $\alpha=h_2$, and we want to solve the spectral problem 
\be
\label{sp-pt}
\mH |\phi\rangle= E |\phi\rangle, \qquad E=\ri h_3.
\ee
The spectrum is expected to have a structure qualitatively similar to the one obtained in \cite{delabaere-trinh} for (\ref{dt-ham}), and 
indeed our analysis will confirm this. We will rely on the EQC conjectured in \cite{gm-deformed}. The comparison of our explicit 
numerical results with the predictions of this EQC will provide additional support for the conjecture of \cite{gm-deformed}, and therefore for the TS/ST correspondence of \cite{ghm,cgm}. 

Before presenting the conjecture of \cite{gm-deformed}, let us make a cautionary remark concerning its general applicability in the study of PT-symmetric Hamiltonians. 
In defining the spectral problem for these Hamiltonians, one has to make a careful choice of boundary conditions for the eigenfunctions. In the case of the 
PT symmetric {\it cubic} oscillator, the boundary condition is square integrability along the 
real axis. This agrees with the boundary condition that one obtains by analytic continuation in the coupling constant
\cite{alvarezcubicbw}. However, there are cases in which the boundary conditions appropriate for PT symmetry 
are different from the boundary conditions obtained by analytic continuation. For example, the quartic oscillator with potential $-x^4$, which is PT symmetric, 
leads to a real spectrum when appropriate boundary conditions are used. If one does instead an analytic continuation in the coupling constant, the appropriate boundary condition is 
exponential decay in the wedge (\ref{rot-wedge}), and one obtains a spectrum of complex resonances \cite{bender-review}. The EQCs obtained in \cite{gm-deformed} describe resonances 
obtained by analytic continuation in the parameters. Therefore, they cannot be used to describe e.g. the real spectrum of an inverted, PT-symmetric quartic oscillator. 

Let us now present the EQC for the Hamiltonian (\ref{def-pt}). By setting $\Lambda=\ri$ in the EQC for $N=3$ in \cite{gm-deformed}, 
one obtains
\be\label{eqcPT}
1
+\frac{1-\re^{-2\pi a_1/\hbar}}{1-\re^{-2\pi (a_1+a_2)/\hbar}}\re^{-\pi a_2/\hbar+\ri\phi_2}
+\frac{1-\re^{-2\pi a_2/\hbar}}{1-\re^{-2\pi (a_1+a_2)/\hbar}}\re^{-\pi a_1/\hbar+\ri\phi_1}
=0,
\ee
where
\be
\phi_1=\frac{1}{\hbar}\left(\frac{\de F_{\rm NS}}{\de a_2}-2\frac{\de F_{\rm NS}}{\de a_1}\right), 
\qquad
\phi_2=\frac{1}{\hbar}\left(2\frac{\de F_{\rm NS}}{\de a_2}-\frac{\de F_{\rm NS}}{\de a_1}\right). 
\ee
When $N=3$, the perturbative piece of the NS free energy is determined by
\be\label{dFNS}
\frac{\de F^{\rm pert}_{\rm NS}}{\de a_j}=2\gamma(a_j,\hbar)+2\gamma(a_1+a_2,\hbar),
\qquad
j=1,2.
\ee
where the function $\gamma(a,\hbar)$ was introduced in \cite{hm} and it is given by\footnote{The definition of $\gamma$ differs from the one in \cite{gm-deformed} by the factor $a\log\Lambda$
which in our case ($\Lambda=\ri$) is already included in \eqref{eqcPT}.}
\be
\gamma(a,\hbar)=a\log\hbar-\frac{\pi\hbar}{4}-\frac{i\hbar}{2}\log\frac{\Gamma(1+\ri a/\hbar)}{\Gamma(1-\ri a/\hbar)}.
\ee
The quantum mirror maps (\ref{hk-gauge}) relating $\alpha,E$ to the quantum periods $a_1,a_2$ are given, at the very first orders in $\Lambda^6$, by
\begin{align}
&\begin{aligned}
\alpha=
&-\frac{1}{3}(a_1^2+a_1a_2+a_2^2) \\
&+\frac{2 (a_1^2+a_2^2+a_1 a_2+3 \hbar ^2)}{(\hbar^2 + a^2_1) (\hbar^2 +a^2_2) 
\left(\hbar^2+(a_1+a_2)^2\right)}(-\Lambda^6)
+\CO(\Lambda^{12}), \label{alpha}
\end{aligned}
\\
&\begin{aligned}
E=
&-\frac{\ri}{27}(a_1-a_2)(a_1+2a_2)(2a_1+a_2) \\
&-\frac{\ri(4 a_1^3+6 a_2 a_1^2-6 a_2^2 a_1-4 a_2^3)}{3(\hbar^2 + a^2_1) (\hbar^2 +a^2_2) 
\left(\hbar^2+(a_1+a_2)^2\right)}(-\Lambda^6)
+\CO(\Lambda^{12}). \label{energy}
\end{aligned}
\end{align}
Although we have written down explicitly the powers of $\Lambda$ to make clear the 
instanton order, we have to set $\Lambda=\ri$ in actual 
computations. 

It will be useful to write down the EQC (\ref{eqcPT}) in a form similar to (\ref{eqc-pt}): 
\be\label{eqcshort}
F(U,Y)=1+2 \re^U \cos Y=0,
\ee
where the functions $U$, $Y$ are given by
\begin{align}
U&=\frac{1}{2}\log\left(\frac{(1-\re^{-2\pi a_1/\hbar})(1-\re^{-2\pi a_2/\hbar})}{(1-\re^{-2\pi (a_1+a_2)/\hbar})^2}\right)
-\frac{\pi}{2\hbar}(a_1+a_2)
+\frac{3\ri}{2\hbar}\left(\frac{\de F_{\rm NS}}{\de a_2}-\frac{\de F_{\rm NS}}{\de a_1}\right), \label{U1} \\
Y&=\frac{1}{2\ri}\log\left(\frac{1-\re^{-2\pi a_1/\hbar}}{1-\re^{-2\pi a_2/\hbar}}\right)
+\frac{\pi}{2 \ri\hbar}(a_1-a_2)
+\frac{1}{2\hbar}\left(\frac{\de F_{\rm NS}}{\de a_2}+\frac{\de F_{\rm NS}}{\de a_1}\right). \label{Y1}
\end{align}
The EQC (\ref{eqcshort}) is a prediction of the TS/ST correspondence for the spectral problem defined by the 
Hamiltonian (\ref{def-pt}). A first question to ask is how this EQC compares to standard 
semiclassical results, like for example the BS quantization condition. If we write the SW curve (\ref{sw-curve}) in the form 
\be
2 \cosh(p)+ V(x) = E, 
\ee
we notice that the turning points are the solutions of the equation 
\be
E-V(x)=2.  
\ee
In the case
\be
V(x)= \ri x^3 + \ri \alpha x, 
\ee
and for $\alpha$ smaller than a certain $\alpha_c>0$, there are three turning 
points $x_0$, $x_\pm$ in the imaginary axis and the fourth and third quadrant, respectively, as in the undeformed case 
considered in (\ref{pot-alpha}). As in \cite{bb-pt, bender-review}, the relevant period 
corresponds to a cycle going from $x_-$ to $x_+$, given by 
\be
\label{pibs}
\Pi_{\rm BS}= 2 \int_{x_-}^{x_+} p(x) \rd x, 
\ee
and the BS quantization condition is 
\be
\label{bs-qc}
\Pi_{\rm BS}(E)= \pi \hbar \left( n+{1\over 2} \right). 
\ee
In contrast to the situation in conventional quantum mechanics, this can not be expressed in 
terms of elliptic integrals, since the underlying SW curve has genus two. One possibility to 
calculate the period above is to write down an appropriate Picard--Fuchs operator which annihilates $\Pi_{\rm BS}$, and 
then use it to obtain a power series expansion for large energies. The details are presented in Appendix \ref{pf-sec}, and one 
obtains at the end of the day an expression of the form 
\be
\label{eq:finalPi}
\ba
\Pi_{\rm BS}(E)&=  \frac{1}{2} \epsilon  \left(6 \sqrt{3} \log (\epsilon )+\pi +3 \sqrt{3} (\log (3)-2)\right)+\frac{\alpha  \left(-6 \sqrt{3} \log (\epsilon )+\pi -3 \sqrt{3} \log
	(3)\right)}{6 \epsilon } \\
&+\frac{-6 \sqrt{3} \left(\alpha ^3+18\right) \log (\epsilon )-\pi  \left(\alpha ^3+18\right)-3 \sqrt{3} \left(\alpha ^3 (\log
	(3)-1)-6+18 \log (3)\right)}{162 \epsilon ^5}\\
	&+ \CO(\epsilon^{-7}), 
\ea\ee
where $\epsilon=E^{1/3}$. One can verify that (\ref{bs-qc}) provides an excellent approximation to the energy levels of the PT-symmetric Hamiltonian (\ref{def-pt}), in particular for highly excited states, as expected from the 
BS approximation. 

Let us now compare the BS approximation with the EQC (\ref{eqcshort}). First of all, let us note that the functions $U$ and $Y$ have an asymptotic expansion in powers of $\hbar^2$ of the form
\be
\label{uy-semi}
Y\sim \sum_{n\ge 0} Y_n \hbar^{2n-1}, \qquad  U\sim \sum_{n\ge 0} U_n \hbar^{2n-1}. 
\ee
Note that the logarithmic terms in (\ref{U1}), (\ref{Y1}) do not contribute to this asymptotic expansion, as they are exponentially small in $\hbar$. The asymptotic series (\ref{uy-semi}) can be obtained by using the gauge theory instanton expansion for $F^{\rm inst}_{\rm NS}$ and expanding each order in $\Lambda^6$ in powers of $\hbar^2$. This leads to explicit expressions for $Y_n$, $U_n$ as expansions in $\Lambda^6$, which are suitable for the large radius region in which $|a_1|, \, |a_2| \gg 1$. Let us give an explicit example of such an expansion. To do this, we parametrize 
$a_{1,2}$ as 
\be
\label{polar}
a_1= \rho \,  \re^{\ri\phi},\qquad a_2=\rho \, \re^{-\ri\phi}. 
\ee
When $\rho, \phi \in \IR$, $a_1$ and $a_2$ are complex conjugate
\be
a_1=\overline{a_2},
\ee
and this leads to real values of $\alpha$ and $E$, so this describes the phase of unbroken PT symmetry. 
In addition, $U$ and $Y$ are also real. The gauge theory instanton expansion of the function $Y_0$ is, in the polar variables above, 
\be
\ba
Y_0
&=\rho \big((\pi-2\phi)\sin\phi+(-6+6\log\rho+4\log(2\cos\phi))\cos\phi\big) \\
&-\frac{(7 \cos (2 \phi )+2 \cos (4 \phi )+3) \sec ^3(\phi )}{4 \rho ^5  }(-\Lambda^6)+\CO\left(\Lambda^{12}\right), \label{Yh}
\ea
\ee
while for the function $U_0$ one finds, 
\be
 \label{Uh}
\ba
U_0&=\rho \big((6\phi-\pi)\cos\phi+(-6+6\log\rho)\sin\phi \big)
+\frac{3 (4 \cos (2 \phi )+3) \tan (\phi ) \sec (\phi )}{2 \rho ^5 }(-\Lambda^6) \\
&+\CO\left( \Lambda^{12}\right).
 \ea
\ee
Note that the power of $\Lambda$ is correlated with the inverse power of $\rho$, 
so the expansion in powers of $\Lambda$, typical of the gauge theory instanton expansion, is also an expansion around the large radius 
region $\rho\rightarrow \infty$, as expected.

The BS quantization condition should be recovered from (\ref{eqcshort}) by setting 
$\re^{-U} \sim 0$ and keeping the first term in the asymptotic expansion of $Y$, i.e.\ 
\be
\cos(Y_0/\hbar)=0 \quad\Rightarrow\quad Y_0= \pi \hbar \left(n+{1\over 2} \right), 
\ee
so we should have 
\be
\label{recBS}
Y_0=\Pi_{\rm BS}.
\ee
 This can be verified by using the quantum mirror map and expanding $\Pi_{\rm BS}$ in power series around $\rho=\infty$ and $\phi=\pi/3$ 
 (which corresponds to $\alpha=0$). This expansion can be easily compared to the series (\ref{Yh}), and one finds that (\ref{recBS}) holds to very high order.

In view of the structure of the EQC (\ref{eqcshort}), which is formally identical to the one 
in the undeformed case (\ref{eqc-pt}), the deformed Hamiltonian (\ref{def-pt}) might also undergo PT symmetry breaking. 
As in the undeformed case, this can happen if the non-perturbative correction $\re^{-U}$ becomes of order one. 
Semiclassically, this occurs when 
\be
\label{u0-phase}
U_0(\alpha, E)=0. 
\ee
The solutions to this equation, if they exist, define a curve in the $(\alpha, E)$ plane separating the two phases. Note that $U_0$ is completely determined by the explicit expression (\ref{U1}), which is in turn a prediction of the TS/ST correspondence.

 \begin{center}
 \begin{figure} \begin{center}
 {\includegraphics[scale=0.50]{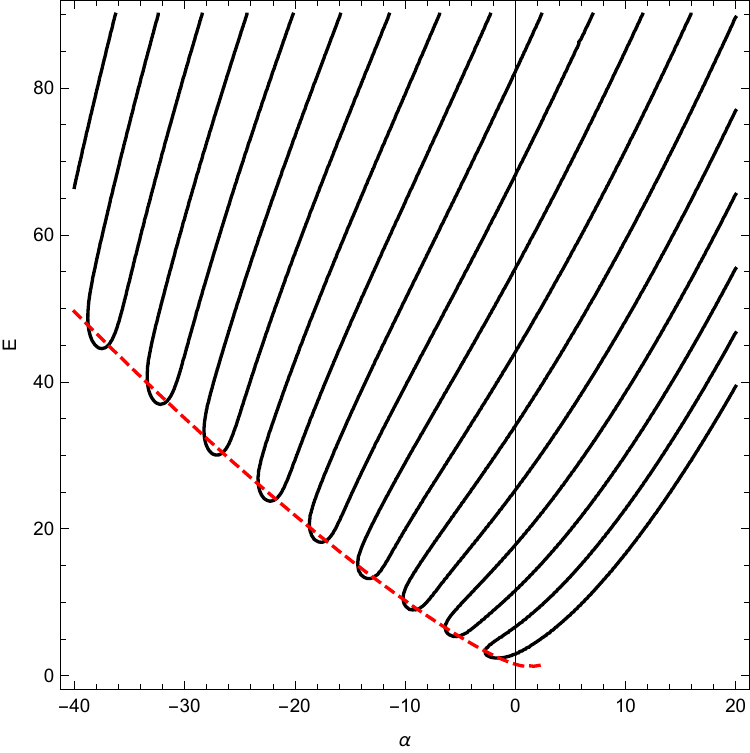}}
  \caption{The black lines show the values of $(\alpha,E)$ in the spectrum of (\ref{def-pt}) at $\hbar=1$. The dashed red line is the solution of the equation (\ref{u0-phase}), which defines the boundary between the phases of broken and unbroken PT symmetry.} \label{BWphases}
    \end{center}
\end{figure}  
\end{center}

\begin{center}
 \begin{figure} \begin{center}
 {\includegraphics[scale=0.7]{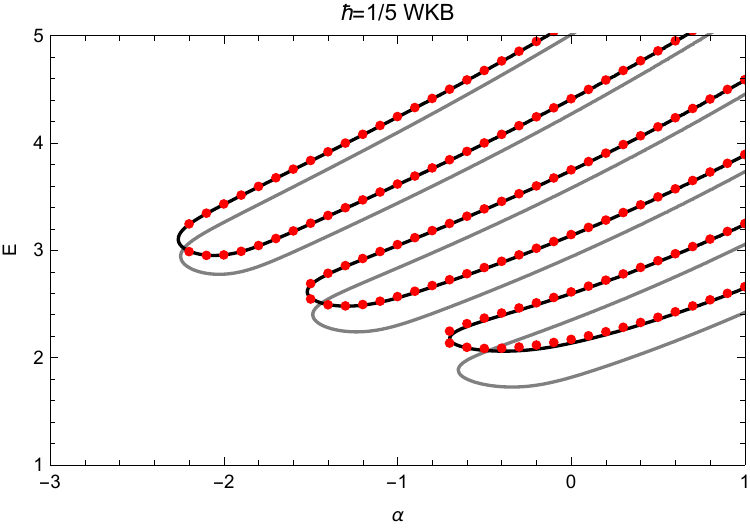}}
  \caption{The black lines show the curves satisfying the approximate EQC (\ref{def-eqc-ap}), in which $U_0$, $Y_0$ are given by the gauge theory instanton expansion up to order $\Lambda^6$. In the gray lines, only the perturbative part  (i.e.\ the very first terms in the expansions (\ref{Yh}), (\ref{Uh})) has been used. The red dots represent numerical values of the spectrum of the operator (\ref{def-pt}), obtained with 
  complex dilatation techniques. Here $\hbar=1/5$.}   
  \label{eqcap-fig}
    \end{center}
\end{figure}  
\end{center}

 It is easy to verify numerically that indeed, PT symmetry breaking takes place in this model, when $\alpha<0$. As in the undeformed case, there is a sequence of values 
 \be
 \label{alphas}
 \alpha_k, \qquad k=1, 2, \cdots, 
 \ee
 for which the energy levels become degenerate $E_{k-1}=E_k$, and take complex conjugate values afterwards. 
 The curve determined by (\ref{u0-phase}) provides a qualitatively accurate boundary between the resulting 
two phases, as shown in \figref{BWphases}. The EQC (\ref{eqcshort}) turns out to capture as well the precise pattern of symmetry breaking. To see this, we consider the leading order approximation to (\ref{eqcshort}), 
but taking into account the contribution of $U$, which is the responsible for symmetry breaking, also at leading order, as we did in (\ref{approx-qc}). We obtain in this way the approximate quantization condition,
\be
\label{def-eqc-ap}
2 \cos(Y_0/\hbar) + \re^{-U_0/\hbar}=0, 
\ee
similar to the condition (\ref{eqc-pt}) in the undeformed case. The equation (\ref{def-eqc-ap}) 
defines a discrete set of curves in the $(\alpha, E)$ plane, which we show in \figref{eqcap-fig} for $\hbar=1/5$. The functions $Y_0$, $U_0$ are defined by the expansions 
(\ref{Yh}), (\ref{Uh}). In drawing the gray lines, we have just used the very first term in this expansion (which takes into account only the perturbative contribution to $F_{\rm NS}$), 
while in drawing the black lines we have 
included the first gauge theory instanton correction. The red dots are numerical values for the spectrum of the operator (\ref{def-pt}), 
and as we can see they are located with good precision on the curves defined by (\ref{def-eqc-ap}).

\begin{center}
 \begin{figure} \begin{center}
 {\includegraphics[scale=0.55]{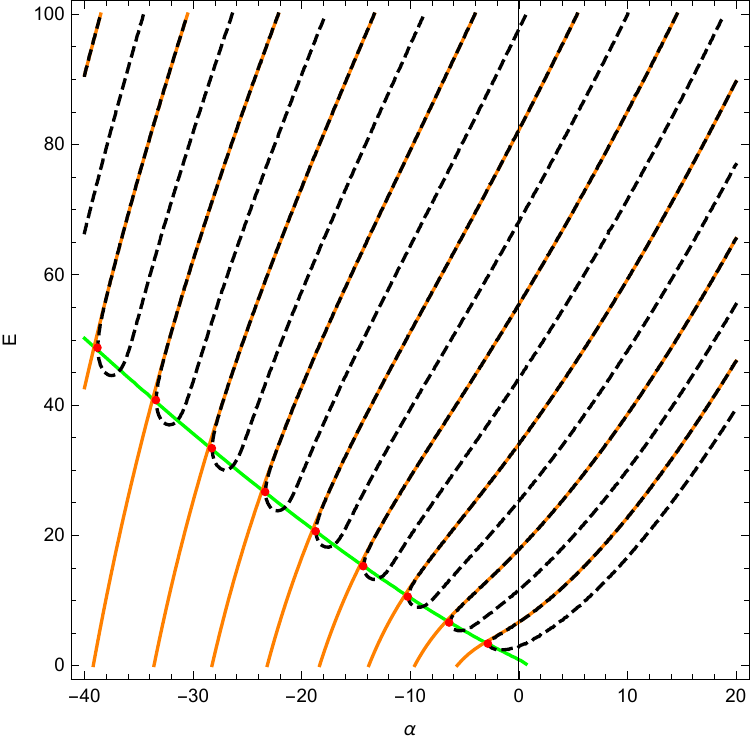}}
  \caption{The black dashed line represents the solution of the spectrum, for $\hbar=1$. The red dots are the branch points.
  The green and orange lines are the curves (\ref{line1}) and (\ref{line2}).}   
  \label{BWpoints-fig}
    \end{center}
\end{figure}  
\end{center}

Another consequence of the EQC is an asymptotic expression for the branch points of the spectrum, i.e.\ for the values of $\alpha$ (\ref{alphas}) at which the energy levels coalesce. These are the points satisfying (\ref{eqcshort}), and in addition the singularity condition 
\be
\label{branch-cond}
{\partial F \over \partial E}=0. 
\ee
Since $\alpha_k$ with $k\gg 1$ occurs at higher energies, in order to obtain the leading asymptotic value of $\alpha_k$ it suffices to use the leading terms in (\ref{Yh}), (\ref{Uh}). A careful 
analysis shows that the branching points obey two different conditions. The first one, not surprisingly, is (\ref{u0-phase}), which is solved by 
\be
\label{line1}
\rho(\phi)\approx \exp\left({\pi \over 6 \phi} \right) 
\ee
The second condition is 
\be
\sin(Y) \approx -1, 
\ee
which is solved by 
\be
\label{line2}
Y\approx 2 k \pi -{ \pi \over 2} , \qquad k=1,2,\cdots. 
\ee
This result is illustrated in \figref{BWpoints-fig}, where we show the curves (\ref{line1}), (\ref{line2}) in the $(\alpha, E)$ plane. Their intersection agrees with good precision with the 
branching points of the spectrum. The values of $\rho$ satisfying the two conditions (\ref{line1}), (\ref{line2}) can be written down explicitly as 
\be
\rho_k=\frac{X_k}{W\left(\re^{\frac{c}{6}} X_k \right)},
\ee
where $W(z)$ is the Lambert $W$ function defined as the solution of the equation $W(z)e^{W(z)}=z$, and
\be
X_k=\frac{\hbar\pi(2k-1/2)}{6},
\qquad
c=4\log 2-6.
\ee

\sectiono{Conclusions and outlook}
\label{conc}

The correspondence between quantum spectral problems and (topological) strings put forward in \cite{ghm,cgm,wzh} has been mostly studied in the 
case in which reality and positivity conditions lead to trace class, self-adjoint operators. However, it is natural to extend this correspondence 
to general complex values of the mass parameters. On the topological string side, the mass parameters are identified with 
complex moduli of the CY target, and we can perform a straightforward analytic continuation. In the spectral theory side, this continuation 
is more delicate since one has to consider a problem in non-Hermitian quantum mechanics. In this paper we have started to explore the TS/ST 
correspondence in the complex realm of the mass parameters. In the spectral theory side, we have developed techniques to 
compute accurately the complex spectra, by using complex dilatation techniques and Borel--Pad\'e resummation of perturbative series. We have found, in various models, that 
these techniques provide results in precise agreement with the predictions of the TS/ST correspondence. 

In addition, we have explored PT-symmetric Hamiltonians arising in the quantization of SW curves. These Hamiltonians provide an interesting 
generalizations of the more conventional PT-symmetric quantum-mechanical models, and they can be analyzed in detail by using the exact quantization conditions 
conjectured in \cite{gm-deformed}, which also followed from the TS/ST correspondence. In this paper we have analyzed in detail the deformed model of \cite{gm-deformed} with a 
PT-symmetric cubic potential. We have shown 
that the quantization conditions conjectured in \cite{gm-deformed} describe very precisely the properties of the spectrum. In addition, we have shown that the 
deformed model displays as well the spontaneous breaking of PT-symmetry 
observed in \cite{delabaere-trinh,ben-ber}. 

There are many interesting problems open by this investigation. The most important one, in our view, is a deeper understanding of the multivalued structure 
of the spectrum of quantum mirror curves. As we have pointed out, the exact solution conjectured in \cite{ghm,cgm,wzh} indicates that there should be an infinitely-sheeted 
structure due to the logarithmic dependence on the mass parameters. Although we have been able to access some of these sheets in spectral theory, a full understanding of this 
structure is still lacking. In addition, the exact quantization conditions in \cite{ghm,cgm,wzh} involve series expansions which do not always converge when one considers arbitrary sheets of the 
logarithm. This raises the possibility that these conditions have to be reformulated in order to describe the full analytic continuation of the spectral problem. 

In our analysis of PT-symmetric models, we have focused for simplicity on quantum SW curves, but it would be nice to find interesting examples of 
PT-symmetric quantum mirror curves\footnote{Spectral problems with PT 
symmetry have been recently considered in \cite{ar-pt}, in the closely related context of the Fermi gas approach to ABJM theories and generalizations thereof.}. 
Conversely, in the case of quantum SW curves, we have not studied in detail the general multivalued structure of the spectral problem. 
Since these models are deformations of conventional 
quantum mechanical potentials, we should have analogues of the Bender--Wu branch points for the quartic \cite{bw} and the cubic case \cite{alvarezcubicbw}. It would be very interesting 
to explore the analytic structure of the energy levels with the exact quantization conditions of \cite{gm-deformed}, similar to what was 
done in \cite{unfolding,delabaere-trinh}. This might lead to a 
beautiful interplay between the complex geometry of the spectral problem and the quantum moduli space of Seiberg--Witten theory. 

\section*{Acknowledgements}

We are grateful to C. Bender and J. Gu for useful discussions. We are particularly grateful to S. Zakany for providing us his 
powerful {\it Mathematica} packages to implement exact quantization conditions. Many of the computations reported 
in this paper were performed at University of Geneva on the Baobab cluster.
The work of Y.E. and M.M. is supported in part by the Fonds 
National Suisse, subsidy 200020-175539, 
and by the Swiss-NSF grant NCCR 51NF40-182902 ``The Mathematics of Physics'' (SwissMAP). 
The work of M.R.~is supported by the funds awarded by the Friuli Venezia Giulia autonomous
Region Operational Program of the European Social Fund 2014/2020,
project ``HEaD - HIGHER EDUCATION AND DEVELOPMENT SISSA OPERAZIONE 3'', CUP G32F16000080009,
by INFN via Iniziativa Specifica GAST and
by National Group of Mathematical Physics (GNFM-INdAM).
M.R.~would like to thank the GGI for hospitality
during the workshop ``Supersymmetric Quantum Field Theories in the Non-perturbative Regime''
and CERN for the hospitality during his stay in Geneva.

\appendix 

\sectiono{Picard--Fuchs equation for the deformed cubic oscillator}
\label{pf-sec}

An efficient way to compute the BS period (\ref{pibs}) is to find out the Picard--Fuchs (PF) equation that it satisfies, and then solve this equation in a power series in appropriate variables. 
As a starting point, one can derive the following PF equation when $\alpha=0$, corresponding to the pure cubic potential:
\be
\label{eq:PFalphaZero}
\Pi '(E)+\frac{3}{4} \left(3 \left(E^2-4\right) \Pi ^{(3)}(E)+7 E~ \Pi ''(E)\right) = 0.
\ee
To simplify notation, we have denoted $\Pi=\Pi_{\rm BS}$. This ODE can be solved in closed form as
\be
\label{eq:PFalphaZeroSol}
\Pi '(E) = c_1 \frac{P\left(-\frac{1}{2},\frac{1}{6},\frac{E}{2}\right)}{\sqrt[12]{E^2-4}}+c_2\frac{
	Q\left(-\frac{1}{2},\frac{1}{6},\frac{E}{2}\right)}{\sqrt[12]{E^2-4}},
\ee
where $P(n,m,x)$ are Legendre polynomials, and $Q(n,m,x)$ are Legendre functions of the second kind. The PF equation for arbitrary $\alpha$ is more involved:
\begin{eqnarray}
	0 & = & 192 \left(8\,\alpha^3+\,E^2-216\right) \Pi'(E)-\left(-3483 \,E^3-18064 \,\alpha^3 \,E+487728 \,E\right) \Pi ''(E)\nonumber\\
	& & -\frac{1}{3} \left(3328 \,\alpha ^6+6912 \,\alpha^3-18063 \,E^4-71088 \,\alpha^3 \,E^2+1942704
	\,E^2-2612736\right) \Pi ^{(3)}(E)\nonumber\\
	& & -\frac{2}{3} \left(-3645 \,E^5-11232 \,\alpha^3 \,E^3+326592 \,E^3+1744 \,\alpha ^6 \,E-864 \,\alpha^3 \,E-1248048 \,E\right) \Pi ^{(4)}(E)\nonumber\\
	& & -\frac{1}{9} (-128 \,\alpha^9+10368 \,\alpha^6-279936 \,\alpha^3-2187 \,E^6-5184 \,\alpha^3 \,E^4+174960 \,E^4+1680 \,\alpha ^6 \,E^2\nonumber\\
	& &~~~~~~+2592 \,\alpha^3 \,E^2-1294704 \,E^2+2519424) \Pi ^{(5)}(E)  \label{eq:PF}
\end{eqnarray}
To solve this ODE, we use the  following ansatz
\be
\label{eq:ansatzB}
\Pi'(E) = \log (E) \sum _{n=0}^\infty a_n E^{-\frac{2 n}{3}-\frac{2}{3}}+\sum _{n=0}^\infty b_n E^{-\frac{2 n}{3}-\frac{2}{3}}
\ee
By plugging (\ref{eq:ansatzB}) into (\ref{eq:PF}), we can solve for $a_n$ and $b_n$. Doing so, we are left with only four undetermined constants, 
say $a_n$ and $b_n$ with $n \in \{0,1\}$. These can be fixed by various boundary conditions for the problem, and they turn out to be given by 
\begin{align}
&a_0 = -\frac{1}{\sqrt{3}},~~~~b_0=\frac{1}{6} \left(-\pi -3 \sqrt{3} \log (3)\right),\\
&a_1 = -\frac{\alpha }{3 \sqrt{3}},~~~~b_1=\frac{\alpha}{18}   \left(\pi -3 \sqrt{3} (\log
(3)-2)\right).\nonumber
\end{align} 
By using the above results, one can derive the expansion (\ref{eq:finalPi}). Let us note that (\ref{pibs}) is a period of $SU(3)$ SW theory, so it should be possible to 
write it as a linear combination of the basis of periods obtained in \cite{klt} in terms of Appell functions.

\bibliographystyle{JHEP}

\linespread{0.6}
\bibliography{biblio-res}

\end{document}